\documentclass[twocolumn]{aa} % for a paper on 1 column  
\usepackage{graphicx}
\usepackage{txfonts}
\usepackage{CJK}
\usepackage{booktabs}
\usepackage{graphbox}
\usepackage{amsmath}
\usepackage{soul}
\usepackage{scrextend}
\deffootnote[0.45em]{0.2em}{0.2em}{\textsuperscript{\thefootnotemark}\,}
\usepackage{amsmath,url}
\usepackage{algorithm}
\usepackage{amsfonts}
\usepackage{mathrsfs}
\usepackage{bm}
\usepackage{rotating}
\usepackage{color}
\usepackage{graphicx}
\usepackage{subfigure}
\usepackage[toc,page]{appendix}
\usepackage[english]{babel}
\usepackage{orcidlink}
\usepackage[mathlines,switch]{lineno}
\usepackage{todonotes}
\usepackage{hyperref}
\hypersetup{colorlinks=true, allcolors=blue}
\newcommand\gaia{\textit{Gaia~}}
\newcommand\gdr[1]{\gaia DR#1}

\newcommand\masyr{\ensuremath{\text{mas~yr}^{-1}}}

\newcommand\kms{\ensuremath{\text{km~s}^{-1}}}

\newcommand\RAdeg{\ensuremath{\text{$\alpha$}}}
\newcommand\DEdeg{\ensuremath{\text{$\delta$}}}
\newcommand\plx{\ensuremath{\text{$\varpi$}}}
\newcommand\pmra{\ensuremath{\text{$\mu_{\alpha}^*$}}}
\newcommand\pmdec{\ensuremath{\text{$\mu_{\delta}$}}}

\newcommand\figref[1]{Fig.~\ref{#1}}
\newcommand\tabref[1]{Table~\ref{#1}}
\newcommand\secref[1]{Section~\ref{#1}}
\newcommand\appref[1]{Appendix~\ref{#1}}
\newcommand\equref[1]{Eq.~\ref{#1}}
\defcitealias{Hunt2024}{HR24}
\defcitealias{Hunt2023}{HR23}
\defcitealias{Swiggum2024Natur.631...49S}{S24}

\titlerunning{}
\authorrunning{Liu et al}
\begin{document} 
% \linenumbers
\begin{CJK*}{UTF8}{gbsn}
   % \title{Hierarchical structure of Galactic open clusters revealed by two-point correlation analysis}
   \title{The fading hierarchy of Galactic open clusters}
   \author{Guimei Liu \inst{1,2,3}
   \and {Jo\~ao Alves}\inst{2}
   \and {Yu Zhang} \inst{1,3}
   \and {Emily L. Hunt}\inst{2}
   \and{Ruqiu Lin}\inst{4}
   \and{Josefa~E.~Gro\ss schedl}\inst{5}
   \and{Efrem Maconi}\inst{2}
   \and{Nora Wagner}\inst{2}
   \and{Lilly Kormann}\inst{2}
   \and{Cameren Swiggum}\inst{6}
   }
   \institute{
   XinJiang Astronomical Observatory, Chinese Academy of Sciences,150 Science 1-Street, Urumqi, Xinjiang 830011, China \email{zhangyu@xao.ac.cn}  
   \and Department of Astrophysics, University of Vienna, T\"urkenschanzstrasse 17, 1180 Wien, Austria\label{univie}
   \and School of Astronomy and Space Science, University of Chinese Academy of Sciences, No. 19A, Yuquan Road, Beijing 100049, China
    \and University of Massachusetts Amherst, 710 North Pleasant Street, Amherst, MA 01003-9305, USA
   \and Astronomical Institute of the Czech Academy of Sciences, Bo\v{c}n\'{i} II 1401, 141 31 Prague 4, Czech Republic
    \and Center for Astrophysics | Harvard \& Smithsonian, 60 Garden Street, Cambridge, MA 02138, USA
   }
   \date{Received ..., 20XX; accepted ...}
% \abstract{}{}{}{}{} 
% 5 {} token are mandatory
\abstract
{Galactic open clusters provide a record of both hierarchical star formation and the subsequent dynamical evolution of the Milky Way disk. We use the two-point correlation function to characterize the spatial and kinematic clustering of open clusters in the solar neighborhood, considering their three-dimensional (3D) distributions and their projections onto the Galactic plane, as well as subsamples divided by age, mass, and survey volume. We quantify the clustering through fractal dimensions and characteristic correlation scales.
We find that the youngest clusters have fractal dimensions comparable to those measured in the interstellar medium, suggesting that they retain part of the hierarchical structure of their natal molecular clouds. The clustering strength then decreases systematically with age and becomes weak on a characteristic timescale of order 100 Myr. This evolution is scale dependent: the excess correlation at tens of parsecs fades more rapidly than the weaker correlation at hundreds of parsecs. By contrast, we find only a weak dependence of the clustering properties on cluster mass and volume. Using the same cluster sample, we derive an empirical mapping between projected and 3D fractal dimensions, providing an observational calibration for comparing the Milky Way with studies of external galaxies, where only projected measurements are available. The open-cluster population exhibits characteristic spatial and kinematic clustering scales of approximately 370 pc and 11~\kms, respectively. 
These results support a picture in which young open clusters inherit a spatial hierarchy from star formation and progressively lose their spatial correlations through subsequent evolution.}
   \maketitle
%===========================================
%introduction
%===========================================
\section{Introduction}\label{sec:intro}
Star formation proceeds in a spatially and temporally correlated manner and commonly produces star clusters \citep{Lada2003}. It is widely recognized as a hierarchical and self-similar process that reflects the fractal structure of the interstellar medium \citep[ISM;][]{Elmegreen2004ARA&A..42..211E,Kruijssen2012}. Under this framework, giant molecular clouds (GMCs) undergo multi-level fragmentation driven by turbulence \citep{Larson_1981MNRAS.194..809L}. This process creates a nested hierarchy of structures that range from dense cores and filaments to star clusters and larger systems known as star cluster complexes \citep{Mac-Low2004, Andre2014prpl}. Consequently, young star clusters typically reside in clustered configurations within star-forming complexes on kiloparsec (kpc) scales \citep{Efremov+1978, Elmegreen2006astro.ph..5519E}. In early evolutionary stages, star cluster complexes retain the primordial imprint of their natal clouds and show significant spatial clustering on scales from sub-parsec to several hundred parsecs.

% progress in OC group
Recent observational studies increasingly demonstrate that open clusters (OCs) do not form or evolve as isolated systems \citep{Piecka21,Ratzenb+2023,Swiggum2024Natur.631...49S,posch2024clusterchains,Liu2025a,Liu2025b}. Instead, they are frequently found within higher-order spatial structures, such as binary clusters and star cluster families, which represent hierarchical star formation on intermediate spatial scales. For example, \citet{Swiggum2024Natur.631...49S} found that 155 young star clusters in the solar neighborhood originated from three distinct and large star-forming complexes, identified as three cluster families. 
Together with the widespread occurrence of binary clusters \citep{Liu2025b}, these results provide direct observational evidence that the spatial distribution of OCs extends hierarchically from individual clusters to cluster families.

Despite these advances, it remains unclear whether star clusters are intrinsically clustered and how strong this clustering is, how it depends on age and mass, what the corresponding dissolution timescale is, and whether clusters can serve as reliable tracers of the hierarchical structure of star formation.
Most statistical characterizations of the clustering hierarchy have been derived from extragalactic studies \citep{Gouliermis2015b_ngc6503,Grasha2015ApJ...815...93G,Grasha2017_5,Menon2021MNRAS.507.5542M}. 
However, such analyzes are limited to the projected 2D distribution, making the characteristic scales of clustering inferred from the 2D analyzes biased by the line-of-sight thickness of the galactic disk \citep{Grasha2015ApJ...815...93G,Menon2021MNRAS.507.5542M}, rather than reflecting the true physical extent of the star complex. 
In addition, the fractal dimension ($D$) derived from 2D space ($D_2$) does not have a unique or straightforward correspondence to the intrinsic 3D fractal dimension ($D_3$) from observation, as the mapping depends on projection effects and the underlying geometry of the system \citep{Gouliermis2014MNRAS_ngc346}.
Consequently, 2D analyzes inevitably mix intrinsic clustering properties with geometric projection effects, limiting their ability to recover the true hierarchical structure of star formation.

A full 3D analysis is therefore required to unravel these effects and can be performed in the Milky Way, where accurate distances and kinematics are available. A few studies have quantified the assembly and clustering properties of the Milky Way using OCs \citep{Alfaro2022ApJ...937..114A} or field stars \citep{Kamdar2021ApJ, Hinkel2023ApJ...942...41H}, revealing significant spatial correlations in the Galactic disk. Compared to field stars, OCs provide more reliable tracer of large-scale structure. 
However, earlier OC-based studies have been limited by the relatively small sizes of available OC samples and by the lack of high-precision measurements of distances and radial velocities. These limitations have prevented robust statistical characterization of the clustering hierarchy and its evolution.

The OC census has expanded significantly in the era of high-precision \gaia astrometry, particularly with \gdr{3} and the application of machine-learning techniques for the determination of membership \citep{Cantat2020_1867,Qin2023ApJS,Hunt2023,Hunt2024}. The availability of large and homogeneous cluster catalogs now enables 3D statistical analyzes of OC clustering properties with unprecedented precision.

In this work, we used a large high-quality OC sample from \citealt{Hunt2024} (hereafter \citetalias{Hunt2024}) to quantify the spatial and kinematic clustering properties of OC within 2.8 kpc. We measure the characteristic clustering scales of OCs in both 2D and 3D and investigate their dependence on age, mass and volume. 
Through this analysis, we aim to place the present-day distribution of OCs within the broader framework of hierarchical star formation and the dynamical evolution of the Galactic disk.

This paper is organized as follows. \secref{sec:data} describes the OC samples, together with the completeness criteria. \secref{sec: method} presents the clustering analysis methodology. \secref{sec:results} reports clustering properties of OCs and the empirical calibration between 2D and 3D fractal dimensions.\secref{sec:discussion} discusses the implications for hierarchical star formation and Galactic stellar structures. Finally, \secref{sec:summary} summarizes the main conclusions.
%===========================================
% data
%===========================================
\section{Data} \label{sec:data}

\subsection{The open cluster sample}\label{data:OCs}
Our analysis is based on the homogeneous star cluster catalog compiled by \citetalias{Hunt2024}. This catalog builds upon the systematic search conducted by \citealt{Hunt2023} (hereafter \citetalias{Hunt2023}). \citetalias{Hunt2024} further refined the catalog and provided updated astrometric and physical parameters, while distinguishing gravitationally bound systems, such as OCs, from dynamically unbound structures compatible with moving groups.

To ensure a reliable sample, \citet{Hunt2021} introduced the statistical cluster significance test (CST), which quantifies the statistical contrast between candidate members and the surrounding field stars. Only OCs with a CST greater than 3$\sigma$ were retained, with a CST cut of 5$\sigma$ being recommended for producing a more high-quality cluster sample. 
In addition, cluster reliability was evaluated through color–magnitude diagram (CMD, \citetalias{Hunt2023}), with median CMD class values above 0.5 were classified as consistent with a single stellar population. 

In this work, we restricted our analysis to clusters located within 2.8 kpc of the Sun, corresponding to the distance range over which the catalog remains highly complete for the mass regime relevant to this study (\citetalias{Hunt2024}). After applying the CST and CMD quality criteria, the final sample contains 2648 OCs. We adopted the astrometric parameters (\RAdeg, \DEdeg,\pmra,\pmdec,\plx) and the radial velocities (RVs) together with all relevant physical parameters, including ages, masses, provided by \citetalias{Hunt2024}.
For the 3D velocity analysis, we used the 2289 OCs with available RV measurements. We also excluded \textit{Theia 3475} due to its anomalous RV of approximately $600$~\kms, which could disproportionately affect the velocity-space statistics.

The resulting sample provides a statistically robust view of the present-day OC population spanning a broad range of ages and masses, making it a particularly ideal dataset for investigating the hierarchical cluster formation and its subsequent dynamical evolution.

\subsection{Age-binned samples}\label{sec:data_age}
To investigate how the clustering properties evolve with age, we divided the sample into six age bins. 
Specifically, OCs with $\log(\mathrm{age/yr}) < 9$ were grouped into five logarithmically-spaced age bins and $\log(\mathrm{age/yr}) \ge 9$ were combined into a single bin. The adopted age bins are summarized in \tabref{tab:age_results}.
%----------------------------------------------------------------------
\subsection{The mass-distance dependent volume cut}\label{sec: complete}

The detectability of star clusters in \gdr{3} is strongly dependent on both cluster mass and distance. Low-mass systems become progressively more difficult to identify at larger heliocentric distances owing to magnitude limit, extinction effects, and astrometric uncertainties (\citetalias{Hunt2024}). Consequently, any comparison of clustering properties across different mass or distance ranges must account for these selection effects.

According to the mass-distance completeness model presented by \citetalias{Hunt2024}, the radius of 100\% completeness ($R_{100\%}$) increases approximately log-linearly with the star cluster mass up to $\sim$1000 $M_\odot$. Above this mass, the relation reaches an upper limit at $R_{\rm break}\simeq$2.8 kpc, suggesting that this distance represents the practical upper limit of Gaia's 100\% completeness for OC detection. 
Neglecting this mass-distance dependence would introduce a bias due to selection incompleteness. Therefore, we only used local clusters within 2.8 kpc for our analysis.
Following this completeness framework, we divide the sample into mass- and distance-binned subsamples for the subsequent TPCF analysis. The adopted mass and distance intervals, together with the corresponding numbers of OCs in each subsample, are summarized in \tabref{tab:mass_dist_results}.

\subsubsection{Mass-binned samples}\label{sec:data_mass}
To investigate the impact of cluster mass on clustering,
we divided the sample into five mass intervals and computed the 3D clustering strength for each mass bin. 
To ensure the statistical reliability of each interval, we implemented a mass-dependent distance cut following the completeness model of \citetalias{Hunt2024}. 

For each mass interval, the completeness limit was determined by the lowest mass OCs in the interval.
Although more massive OCs remain detectable farther away, adopting a larger distance limit would introduce an incompleteness bias because the lower-mass OCs would become progressively missing. We therefore adopted the completeness distance of the lowest-mass clusters as the distance limit for the entire mass bin, ensuring a complete and unbiased sample.

\subsubsection{Volume-binned samples}\label{sec:data_dist}
Volume effects within the solar neighborhood are investigated by subdividing the OC sample into five concentric annuli of equal-volume shells based on distance. 
As an illustration, the mass threshold required for full completeness increases with distance. Therefore, for a given distance shell, we adopted the completeness limit corresponding to the outer boundary of the shell. This conservative choice ensures that all clusters included in the subsample would be detectable throughout the entire volume and prevents artificial trends caused by the progressive loss of lower-mass systems at larger distances.

%===========================================
% Method
%===========================================
\section{Methods} \label{sec: method}
\subsection{Coordinate and velocity transformation}

We derived the parameters used in the analysis, including the 3D positions, 3D velocities, and tangential velocities ($V_t$) for all star clusters. 
For high-quality OCs, we converted the equatorial coordinates to Galactocentric Cartesian coordinates using the \texttt{Astropy} package \citep{Astropy2022} based on the 6D astrometric parameters provided by \citetalias{Hunt2024}. We adopted the default solar parameter with Galactocentric radius of 8.122 kpc \citep{Gravity2018}, and a solar height of 20.8 pc \citep{Bennett2019} and velocity of $(U_{\odot},V_{\odot},W_{\odot})=(12.9, 245.6, 7.78)$~\kms \citep{Astropy2022}.

The value of $V_t$ was calculated by converting proper motion measurements as $V_t = k \times d \times PM,$ where $V_t$ is the tangential velocity in \kms, $k=4.74$ is the conversion factor to ensure unit consistency, PM is the proper motion component in \masyr, and $d$ is the heliocentric distance of the clusters (in kpc).

\subsection{Two-point correlation function} \label{sec: method_correlation}
 
The two-point correlation function (TPCF) is a powerful tool to probe the clustering distribution by quantifying the excess probability of finding pairs at a given separation relative to a random Poisson distribution \citep{Peebles1980,LandySzalay1993}. Originally developed for cosmological large-scale structure studies, the TPCF has increasingly been applied to Galactic stellar populations to investigate spatial and kinematic clustering \citep[e.g.][]{Cooper2011MNRAS.417.2206C,Mao2015arXiv150701593M,Lancaster2019MNRAS.484.2556L,Kamdar2021ApJ,Hinkel2023ApJ...942...41H,j.groth1977statistical}.

\begin{figure}
\centering
    \includegraphics[width=0.98\linewidth]{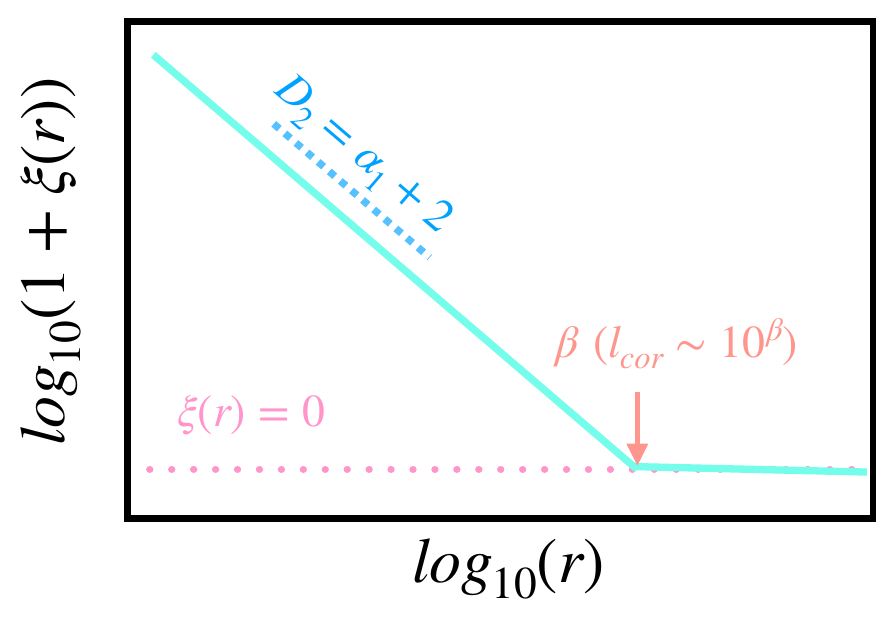}
    \caption{Schematic illustration of the physical parameters of two-point correlation function inferred from the double part power law.}
    \label{fig: schematic}
\end{figure}

To quantify the clustering of OCs on different spatial scales, we calculated the TPCF using the \texttt{astroML} Python package.
In this work, the TPCF was computed both in the projected 2D Galactic plane ($X,Y$) and in 3D Cartesian space ($X,Y,Z$). The Poissonian excess probability of finding two objects separated by a distance $r$, relative to a random distribution, is defined as follows \citep{Peebles1980}:
\begin{equation}
    dP = \bar{n}^{2}\,[1 + \xi(r)]\, dA_{1}\, dA_2,
\end{equation}
where $\bar{n}$ is the mean volume density in 3D or mean surface density in 2D, and $dA$ denotes the corresponding infinitesimal volume element in 3D or area element in 2D. Here, $\xi(r)$ is TPCF, where $\xi(r)=0$ corresponds to a purely random distribution and $\xi(r)>0$ indicates clustering (as shown in \figref{fig: schematic}).

By counting the number of pairs at different separations in the observed sample and comparing them with a normalized random catalog, we calculated the TPCF using the Landy--Szalay estimator \citep{LandySzalay1993}:
\begin{equation}
    \xi(r) =
    \frac{DD(r) - 2\,DR(r) + RR(r)}{RR(r)},
\end{equation}
where $DD(r)$ and $RR(r)$ are the normalized numbers of data-data and random-random pairs at separation $r$, and $DR(r)$ is the normalized cross count of data–-random pairs. These quantities are computed as
\begin{equation}
\begin{aligned}
    DD(r) &= \frac{N_{\mathrm{DD}}}{N_D(N_D-1)/2}, \\
    DR(r) &= \frac{N_{\mathrm{DR}}}{N_D N_{R}}, \\
    RR(r) &= \frac{N_{\mathrm{RR}}}{N_{R}(N_{R}-1)/2},
\end{aligned}
\end{equation}
And the TPCF is assumed to follow a power-law form,
\begin{equation}
    1 + \xi(r) = B_{r}\,r^{\alpha},
\end{equation}
where $\alpha$ characterizes the clustering strength and $B_r$ is the amplitude.
To determine the slope $\alpha$, we fitted the logarithmic form of $1+\xi(r)$ using a linear least-squares method. Hierarchical systems frequently exhibit scale-dependent clustering behavior, producing a break in the TPCF that separates the strongly clustered regime from larger scales where the distribution approaches randomness.
Thus, we adopted the following piecewise model:
\begin{equation}
\log[1+\xi(r)] =
\begin{cases}
    B_{1} + \alpha_{1}\log_{10}(r), 
    & \log(r) < \beta, \\[4pt]
    B_{2} +  \alpha_{2}\log_{10}(r), 
    & \log(r) \geq \beta,
\end{cases}
\end{equation}
where $\beta$ is the break position in the logarithm, 
$B_{1}$ and $B_{2}$ represent the amplitudes on either side of the break,  
and $\alpha_{1}$ and $\alpha_{2}$ are the corresponding inner and outer power-law slopes.  
Among them, the correlation length ($l_{\rm cor}$) is taken to be equivalent to $10^\beta$ \citep{Menon2021MNRAS.507.5542M}, which represents the largest spatial scale over which the fractal clustering signatures are preserved. On scales larger than $l_{\rm cor}$, the spatial distribution of star clusters becomes statistically consistent with a random distribution. 
The quantity of $l_{\rm cor}$ provides a model-dependent estimate of the scale at which the fitted TPCF changes behavior. 
The adopted broken power-law model is illustrated schematically in \figref{fig: schematic}, highlighting the small and large clustering regimes and the characteristic break scale associated with $l_{\rm cor}$. 

%-------------------------------------------------------------------
\subsection{TPCF fitting and uncertainty estimation}
To quantify the sensitivity of the TPCF to the construction of the random reference catalog, we repeated the entire TPCF calculation 100 times. For each separation bin, the median value of the resulting TPCF measurements was adopted as the final estimate. This procedure reduces the sensitivity to individual random realizations and provides a robust characterization of the underlying clustering signal.

For the resulting datasets, the correlation function was transformed into logarithmic space as
\begin{equation}
\Psi_{obs} = \log_{10}(\xi + 1), \quad \chi = \log_{10}(r),
\end{equation}
where $\Psi_{obs}$ represents the observed logarithmic correlation strength and $\chi$ denotes the logarithmic spatial scale. 

We adopted a Markov Chain Monte Carlo (MCMC) method to infer the model parameters using a continuous broken power-law model of the form 
\begin{equation}
\mathcal{M}(\theta|\chi_i)=
\begin{cases}
B_1+\alpha_1\chi,
&
\chi < \beta,
\\[4pt]
B_1+\alpha_1\beta+\alpha_2(\chi-\beta),
&
\chi \geq \beta.
\end{cases}
\end{equation}

The parameter set $\theta$ consists of the break point ($\beta$), the slopes of the two segments ($\alpha_1$ and $\alpha_2$), and the intercept $B_1$. To account for possible model underfitting or additional systematic uncertainties not captured by the nominal measurement errors, we introduced an additional scaling parameter $\ln f$.
The likelihood function was assumed to follow Gaussian statistics and was defined as
\begin{equation}
\ln \mathcal{L}(\theta) = -\frac{1}{2} \sum_i \left[ \frac{(\Psi_{obs,i} - \mathcal{M}(\theta|\chi))^2}{\sigma_{i}^2} + \ln(2\pi \sigma_{i}^2) \right]
\end{equation}
where $\sigma_i^2 = \sigma_{y,i}^2 + \mathcal{M}(\theta|\chi)^2 e^{2\log f}$.
% where \(\sigma_i^2 = \sigma_{y,i}^2 + y_{\rm model}^2 e^{2\log f}\).
The corresponding uncertainty in each separation bin was estimated from the dispersion among the 100 independent TPCF measurements,
\begin{equation}
\sigma_{y,i} = \frac{\sigma_{\Psi_{obs}} + \epsilon}{\sqrt{N_i}}
\end{equation}
where $\sigma_{\Psi_{obs}}$ is the standard deviation of $\Psi_{obs}$ within the bin, and $N_i$ is the number of valid random realizations. The parameter $\epsilon$ represents a small noise floor introduced to avoid underestimating the uncertainty when the number of samples is small or when the standard deviation approaches zero. It was set to $1\%$ of the median standard deviation across all bins. This formulation naturally assigns larger uncertainties to bins with fewer valid realizations, reflecting their lower statistical reliability.
The best-fitting solution obtained from a least-squares fit was used to initialize the MCMC walkers.
The sampling was carried out with the \texttt{emcee} Python package \citep{emcee_2013PASP..125..306F}. We used $32$ walkers, each running for $2000$ steps, and discard the first $500$ steps as burn-in. 
The estimates of the final parameters are taken as the median of the posterior distributions, with the 16th and 84th percentiles adopted as the corresponding uncertainties of $1\sigma$.

\subsection{Scale dependence of clustering dissipation}
To determine whether clustering disappears at the same rate on all spatial scales, we examine how the \emph{shape} of the TPCF changes with age. Fitting a separate decay time at every separation is unreliable when the correlation is weak or fluctuates around zero. Instead, we summarize the relative clustering between compact and larger scales with a single slope for each age cohort:
\begin{equation}
    s(t) \equiv
    \frac{\mathrm{d}\log_{10}[1+\xi(r,t)]}
         {\mathrm{d}\log_{10}r}.
\end{equation}
% We measure this slope over the common range
% $47\leq r\leq632\,\mathrm{pc}$, where the pair counts are sufficiently stable in all four age cohorts. 
A strongly negative value of $s$ means that the correlation is much stronger on compact scales than on larger scales. A value approaching zero means that this compact-scale excess has weakened and the TPCF has become nearly flat. Comparing $s$ between age cohorts therefore provides a direct test of whether compact clustering disappears before correlations on larger scales. 

For each age bin, we recompute the TPCF using a random catalog drawn from an axisymmetric Galactic disk multiplied by the broad empirical selection window described above. This reference population represents the smooth Galactic background and the main survey-selection effects against which the observed clustering is measured. We used five times more random points than observed clusters to reduce the noise from the finite random catalog.

To estimate sampling uncertainties, we divide the observed volume into 12 contiguous regions and repeat the measurement while omitting one region at a time. The variation among these spatial-jackknife measurements provides both the covariance between separation bins and the uncertainty in the fitted slope $s$. We then describe the age dependence of the slope by assigning each cohort a representative age $t$. 
% To match the calculation shown in Fig.~\ref{fig: tpcf_age_mass_dist},
To match the calculation in \secref{sec:data_age},
we use the midpoint of each bin in linear age, giving $t=6.0$, $19.5$, $63.0$, and $202.5\,\mathrm{Myr}$ for the $3$--$9$, $9$--$30$, $30$--$96$, and $96$--$309\,\mathrm{Myr}$ cohorts, respectively. We fit
\begin{equation}
    s(t)=s_0+\eta\log_{10}(t/\mathrm{Myr}).
\end{equation} 
Because the TPCF slopes are initially negative, a positive $\eta$ means that $s$ moves toward zero and the TPCF becomes flatter with age. Physically, this indicates that the excess correlation on compact scales decreases more rapidly than the correlation on larger scales.

%----------------------------------------------------------
\subsection{Fractal dimension}\label{sec:method_fractalD}
The fractal dimension provides a quantitative measure of the hierarchical structure revealed by the TPCF.
For a self-similar 2D distribution, the cumulative number of clusters $N$ scales with the radius as $N \propto r^{D_{2}}$, where $D_{2}$ is the 2D fractal dimension \citep{Mandelbrot1982}.  
Within an annulus of radius $r$, the number of clusters therefore scales as
\begin{equation}
    N(r) \propto r^{\alpha} \times r^{2} \propto r^{\alpha+2}.
\end{equation}
This yields a direct relation $D_{2} = \alpha + 2$ in 2D. Likewise, for 3D distribution, the 3D fractal dimension is given by  $D_3 = 3 + \gamma$ in 3D, where $\alpha$ and $\gamma$ are the slopes of 2D and 3D TPCFs, respectively.
A purely uniform distribution corresponds to $D_{2}=2$ and $D_{3}=3$, whereas smaller $D_2$ or $D_3$ reflect stronger clustering.

%===========================================
% result
%===========================================
\section{Results} \label{sec:results}
\begin{figure*}[htbp]
    \centering
    \includegraphics[width=0.98\linewidth]{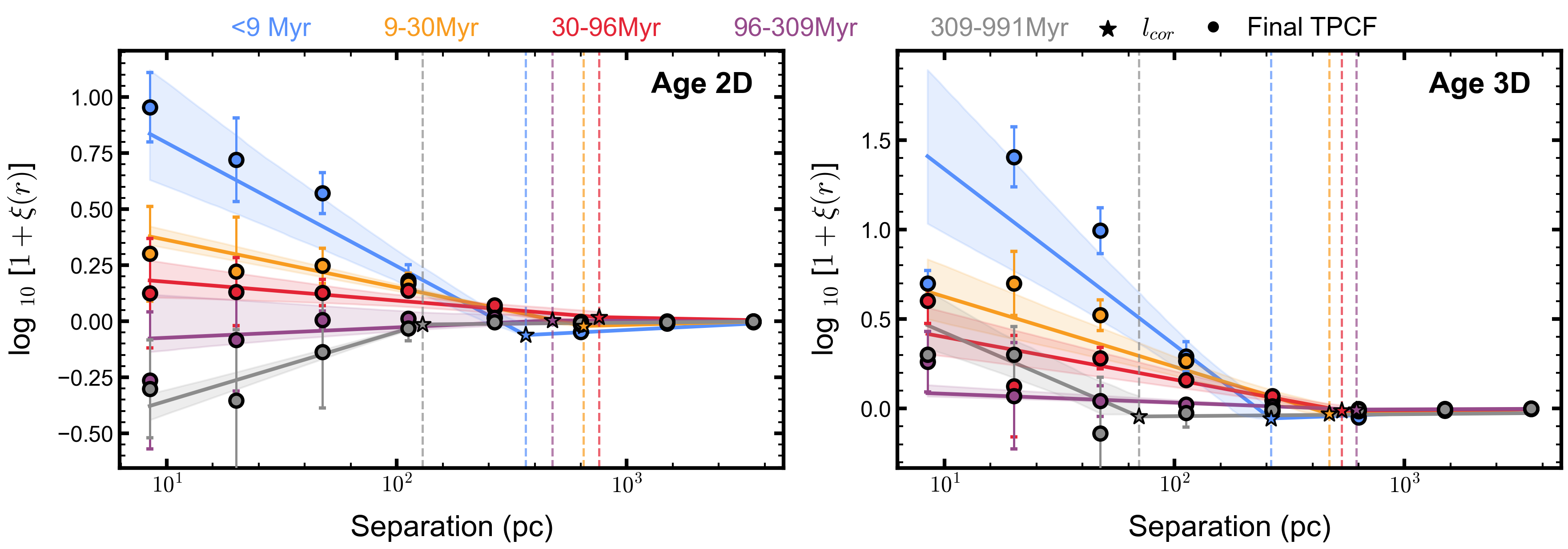}
    \includegraphics[width=0.98\linewidth]{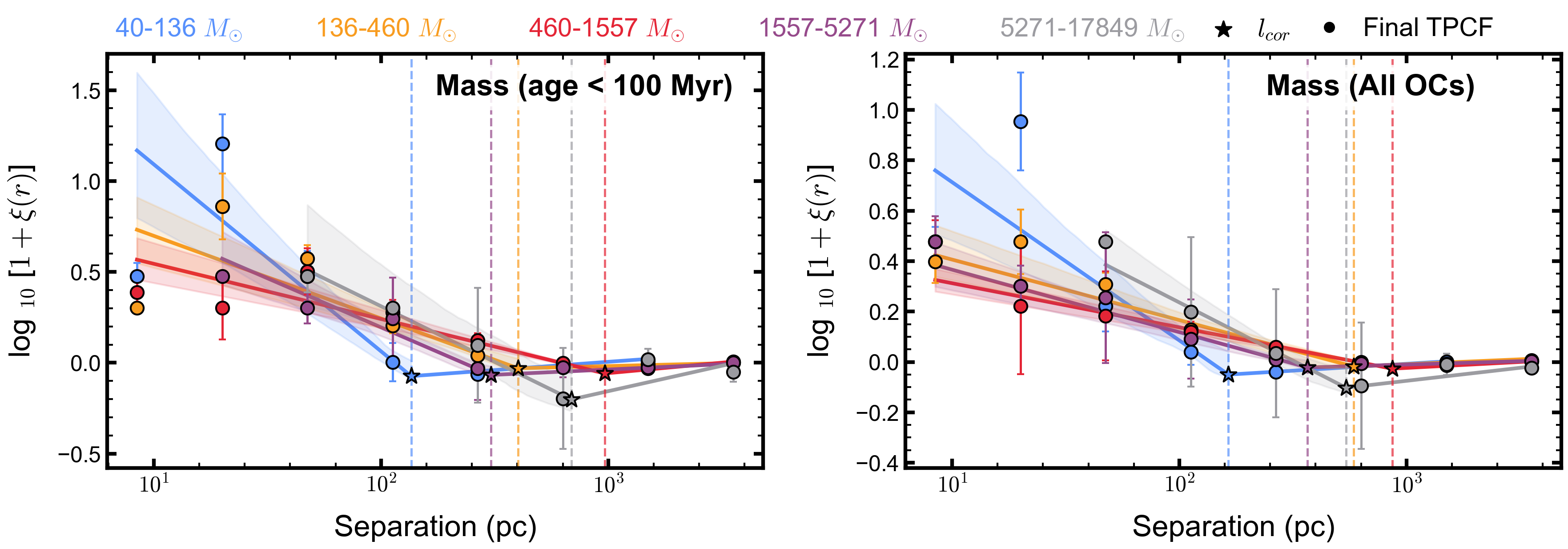}
    \includegraphics[width=0.98\linewidth]{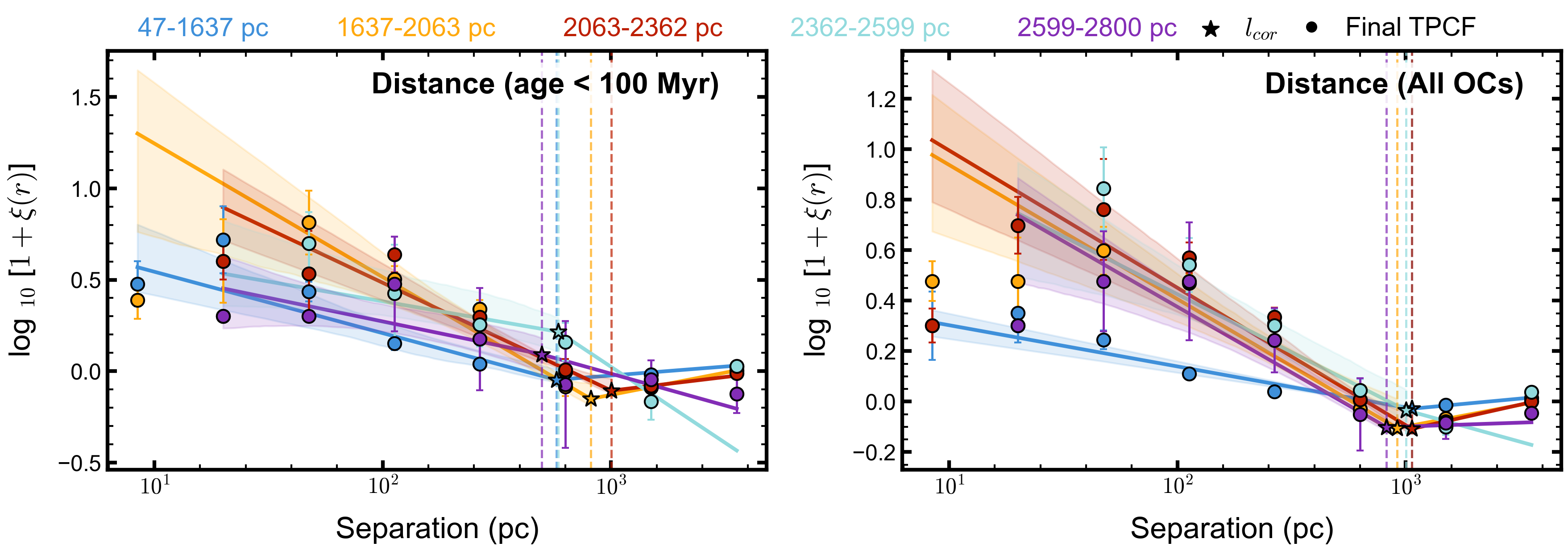}
    \caption{TPCFs ($1+\xi(r)$) for age-binned (top panel), mass-binned (middle panel), and volume-binned sample (bottom panel). Spatial clustering weakens systematically with age, while remaining nearly independent of cluster mass and survey volume.  
    In the top panel, we compared the 2D and 3D TPCFs of age-binned sample. In the middle and bottom panels, we compared age $< 100\,\rm Myr$ subsample with the whole sample for mass-binned and volume-binned samples, respectively.
    Colors denote the different subsamples. Circles represent the median TPCF values from 100 realizations adopted for the broken power-law fits. Dashed vertical lines mark the fitted correlation lengths ($l_{\rm cor}$), which characterize the typical size of OC groups and families.
    }
    \label{fig: tpcf_age_mass_dist}
\end{figure*}

\begin{figure*}
    \includegraphics[width=0.98\linewidth]{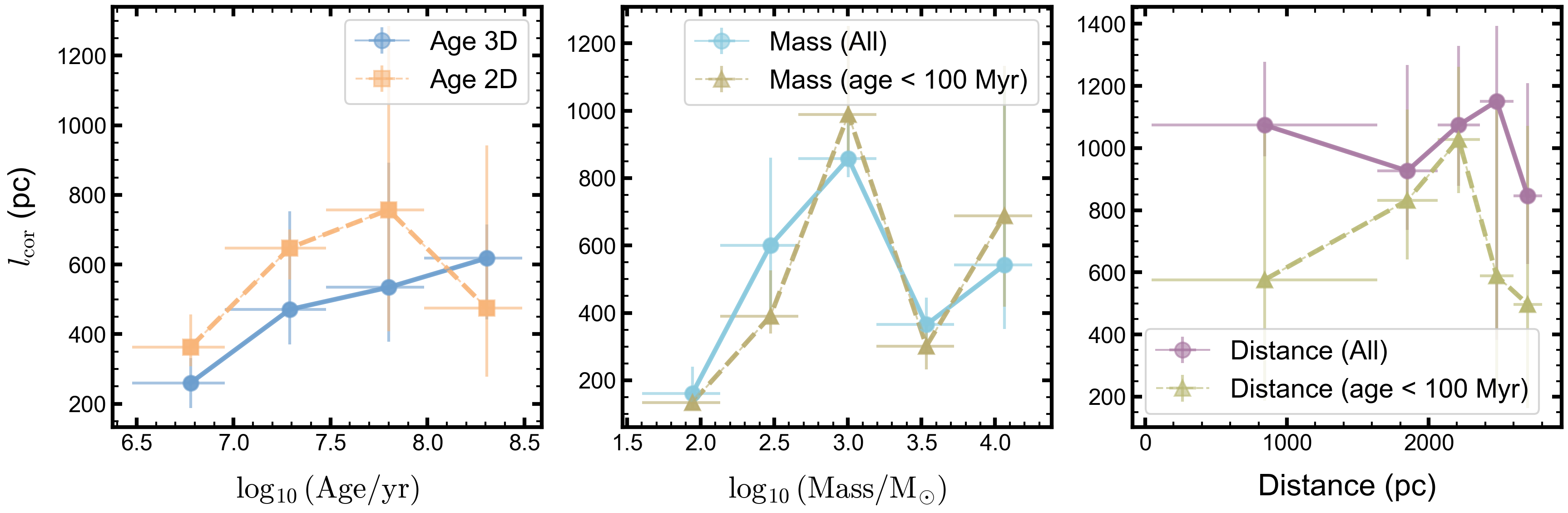}
    \caption{The correlation length ($l_{\rm cor}$) for age-binned, mass-binned, and volume-binned sample. 
    % $l_{\rm cor}$ increase with age in 2D, whereas cluster mass and distance have little influence. 
    \textit{Left:} $l_{\rm cor}$ as a function of age in 2D (orange dashed line) and 3D (blue solid  line). 
    \textit{Middle:} $l_{\rm cor}$ as a function of cluster mass for the full high-quality sample (solid line) and for OCs younger than 100 Myr (dashed line) in 3D space. \textit{Right:} $l_{\rm cor}$ as a function of heliocentric distance for the full sample (solid line) and for OCs younger than 100 Myr (dashed line) in 3D space. Error bars represent the 16th and 84th percentiles of the posterior distributions, corresponding to the $1\sigma$ confidence interval.}
     \label{fig: only_lcor}
\end{figure*}

%---------------- Age effect----------
\subsection{Variation of clustering with age, mass, and volume}

\subsubsection{Clustering as a function of age}\label{sec:res_age}

We computed the TPCF for each age subsample in both 3D space ($X,Y,Z$) and the projected Galactic plane ($X,Y$), with the corresponding fitting results summarized in \tabref{tab:age_results}.
As shown in \figref{fig: tpcf_age_mass_dist}, the slopes of both 2D and 3D TPCFs become progressively flatter with increasing cluster age, indicating a systematic weakening of spatial clustering over time. In the 2D analysis, an obvious transition occurs at a characteristic age of approximately 100 Myr. The 3D TPCF exhibits a similar transition, becoming nearly flat over most spatial scales by the same age. Although the 309--991 Myr bin still shows a weak excess in the two smallest separation bins, the TPCF is already nearly flat over the remaining separation range. As a result, the fitting result is constrained primarily by the first two bins, yielding a small best fitting $l_{\rm cor, 3D}\sim70$ pc. This value is therefore not interpreted as a physical characteristic scale of clustering.
The oldest bin remains poorly constrained due to the small number of clusters and resulting insufficient pair statistics in the small separation bins of the TPCF. 
Therefore, only the first four bins were used in the final analysis. For these four bins, the best-fitting 3D correlation length ($l_{\rm cor,3D}$) generally increases with age, whereas the 2D correlation length ($l_{\rm cor,2D}$) does not show the same systematic trend (\figref{fig: only_lcor}).

\begin{figure*}\centering\includegraphics[width=\textwidth]{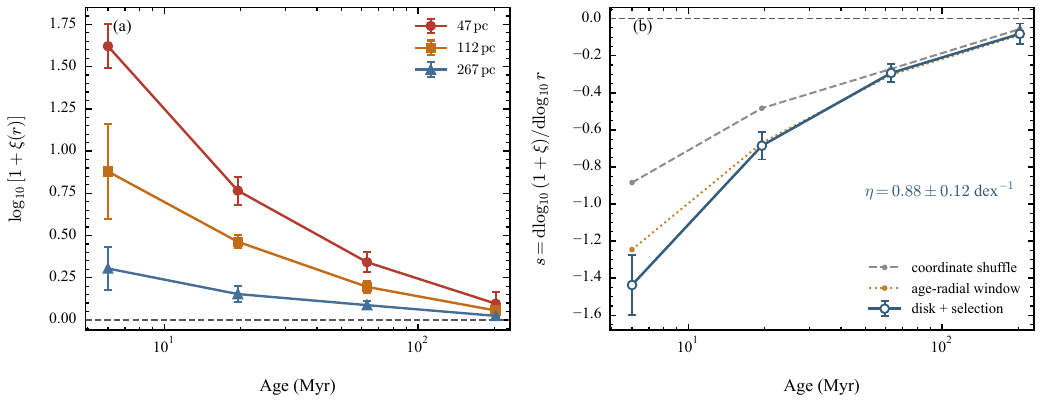}
\caption{Scale-dependent weakening of spatial clustering with age. \textit{Left:} TPCF amplitude at fixed separations. The faster decline at $47\,\mathrm{pc}$ than at $112$ and $267\,\mathrm{pc}$ shows that compact clustering fades first. \textit{Right:} Slope $s$ of the TPCF over $47$--$632\,\mathrm{pc}$. The approach of $s$ toward zero means that the clustering profile becomes progressively flatter with age. Open blue circles show the disk-plus-selection reference with spatial-jackknife uncertainties; the other curves show alternative random-catalog definitions. The positive interaction, $\eta=0.88\pm0.12$ per decade in age, quantifies this scale-dependent evolution.}
\label{fig:scale_dependent_dispersal}
\end{figure*}

As shown in Fig.~\ref{fig:scale_dependent_dispersal}, 
we measure TPCF slope over the common range
$47\leq r\leq632\,\mathrm{pc}$, where the pair counts are sufficiently stable in all four age cohorts. 
The fixed-range slopes change monotonically from
$-1.44\pm0.16$ at $3$--$9\,\mathrm{Myr}$ to
$-0.69\pm0.07$, $-0.29\pm0.05$, and finally
$-0.08\pm0.06$ at $96$--$309\,\mathrm{Myr}$.  The corresponding age--scale
interaction is $\eta=0.88\pm0.12$ per decade in age, where the quoted
uncertainty is the conditional spatial-jackknife error.  The same monotonic
flattening is recovered with the current coordinate-shuffle reference and with
an age-dependent radial-window sensitivity model.  In the oldest of these
cohorts, the $47\,\mathrm{pc}$ excess is consistent with zero
($\xi=0.25\pm0.19$), whereas the correlations at $112$ and $267\,\mathrm{pc}$
remain positive ($\xi=0.14\pm0.06$ and $0.057\pm0.018$, respectively).  These
results support scale-dependent erosion of the clustering pattern: the steep
compact component fades first, while weaker correlations on intermediate
scales remain detectable for longer.  We do not interpret this relative trend
as evidence that clustering power is physically transferred to larger scales.

This behavior suggests that the clustering signal is primarily dominated by young clusters and that the cluster population has transitioned from an initially hierarchical configuration toward a nearly Poisson-like spatial distribution. 
The clustering signal becomes weak beyond ~100 Myr.
Beyond this threshold, the primordial hierarchical structure is effectively erased by dynamical evolution. Therefore, the observed transition near 100 Myr motivates our adoption of this age threshold in the analyses that follow.

%--------------------------
% Mass effect
% \subsubsection{Clustering as a Function of mass}
\subsubsection{Effects of cluster mass and survey volume}
The strong age dependence identified above raises the question of whether the observed evolution of clustering strength could be influenced by other factors, such as cluster mass or observational selection effects associated with heliocentric distance. To assess these possibilities, we examined the 3D TPCF as a function of both cluster mass and survey volume using the completeness-controlled subsamples described in \secref{sec:data_mass} and \secref{sec:data_dist}.

For cluster mass, the analysis was performed in 3D space for both the full high-quality OC samples and a subsample with ages younger than 100 Myr in order to assess whether the mass dependence differs at early evolutionary stages. 
Unlike the strong age dependence described above, the clustering properties show only modest variations across the sampled mass range. The overall similarity of the TPCF slopes indicates that cluster mass is not the primary driver of the observed clustering hierarchy (\figref{fig: tpcf_age_mass_dist}). 
Although $l_{\rm cor,3D}$ varies among different mass bins (\figref{fig: only_lcor}), no robust monotonic trend is established within the uncertainties. Notably, the lowest-mass bin exhibits the steepest slope and a relatively short $l_{\rm cor,3D}$, although the significance of this feature remains limited by the current uncertainties.

%-------------------------- distance
% \subsubsection{Clustering as a Function of Volumes}
A similar result is obtained when the sample is divided according to heliocentric distance. Using the volume-binned subsamples described in \secref{sec:data_dist}, we found that the TPCF slopes remain remarkably similar across most distance intervals for both the young and full samples (\figref{fig: tpcf_age_mass_dist}). No statistically significant monotonic trend with heliocentric distance is detected.
For both the full and young samples, no clear dependence of $l_{\rm cor,3D}$ on heliocentric distance is observed (as shown in \figref{fig: only_lcor}). Overall, the clustering properties of OCs show little variation with heliocentric distance within the sampled volume, suggesting that the underlying hierarchical structure is broadly similar across the solar neighborhood.

Taken together, these results indicate that neither cluster mass nor heliocentric distance produces variations comparable to those observed across the age sequence. The evolution of clustering strength is therefore primarily driven by age, reinforcing the interpretation that the hierarchical structure of OCs is progressively erased through dynamical evolution. 

\subsection{Observed Relation between $D_2$ and $D_3$}\label{sec: calibration}
\begin{figure}
    \centering
    \includegraphics[width=0.98\linewidth]{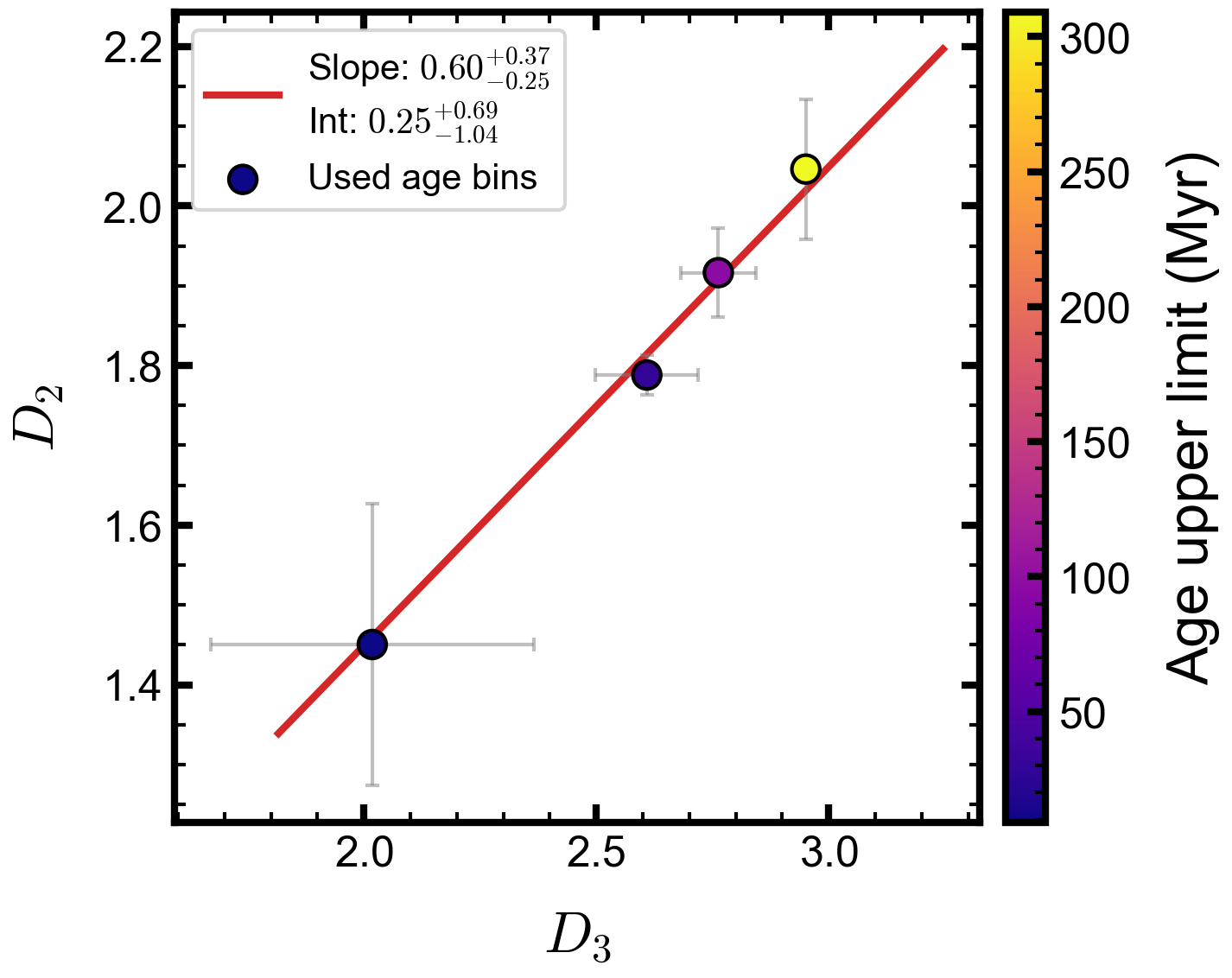}
    \caption{Fractal dimension for different age OC samples in 2D and 3D. Error bars indicate the 16th and 84th percentiles of the distribution obtained from MCMC, corresponding to the 1$\sigma$ confidence interval. The color is coded by the upper edge of each age bin.}
    \label{fig: ageD_cali}
\end{figure}

To establish an empirical relation between the fractal dimension in 2D and 3D in the Milky Way, we used the slope derived from the age-binned TPCF analysis presented in \secref{sec:res_age} to calculate the $D_2$ and $D_3$ values. 
The resulting $D_2$ and $D_3$ values for each age bin are shown in \figref{fig: ageD_cali}, with the associated uncertainties estimated from the MCMC analysis described in \secref{sec: method}. 

We then performed a Monte Carlo sampling to derive the linear relation between $D_2$ and $D_3$. In each realization, $D_2$ and $D_3$ were randomly drawn from Gaussian distributions centered on their measured values, with standard deviations given by their uncertainties. A linear fit was performed for each realization, and this procedure was repeated 10,000 times. The final slope and intercept were taken as the median values of 10,000 realizations, with uncertainties defined by the 16th and 84th percentiles. The resulting empirical relation is 
\begin{equation}
    D_2 = 0.60^{+0.37}_{-0.25} \times D_3 + 0.25^{+0.69}_{-1.04}.
    \label{equ: cal}
\end{equation}
We emphasize that this relation is empirical and was derived specifically from the limited age bin considered here.
In the remainder of this work, this empirical relation was used to convert $D_3$ into projected fractal dimensions $D_2$ scale, enabling a direct comparison of clustering strength between 2D and 3D.

%-------------------------- spatial
\subsection{Spatial clustering in the solar neighborhood}
%fig_output_path
To characterize the global clustering properties of the Galactic OC population, we computed the TPCF for the full high-quality sample of 2648 OCs within 2.8 kpc in both 3D space ($X,Y,Z$) and the projected 2D plane ($X,Y$).
The 3D TPCF traces the intrinsic clustering in real space and includes the vertical structure of the Galactic disk. In contrast, the 2D analysis suppresses variations along the $Z$ axis. A comparison between these two approaches provides an assessment of how vertical structure and projection effects influence the measured clustering signal.

\begin{figure*}
    \centering    
    \includegraphics[width=0.48\linewidth]{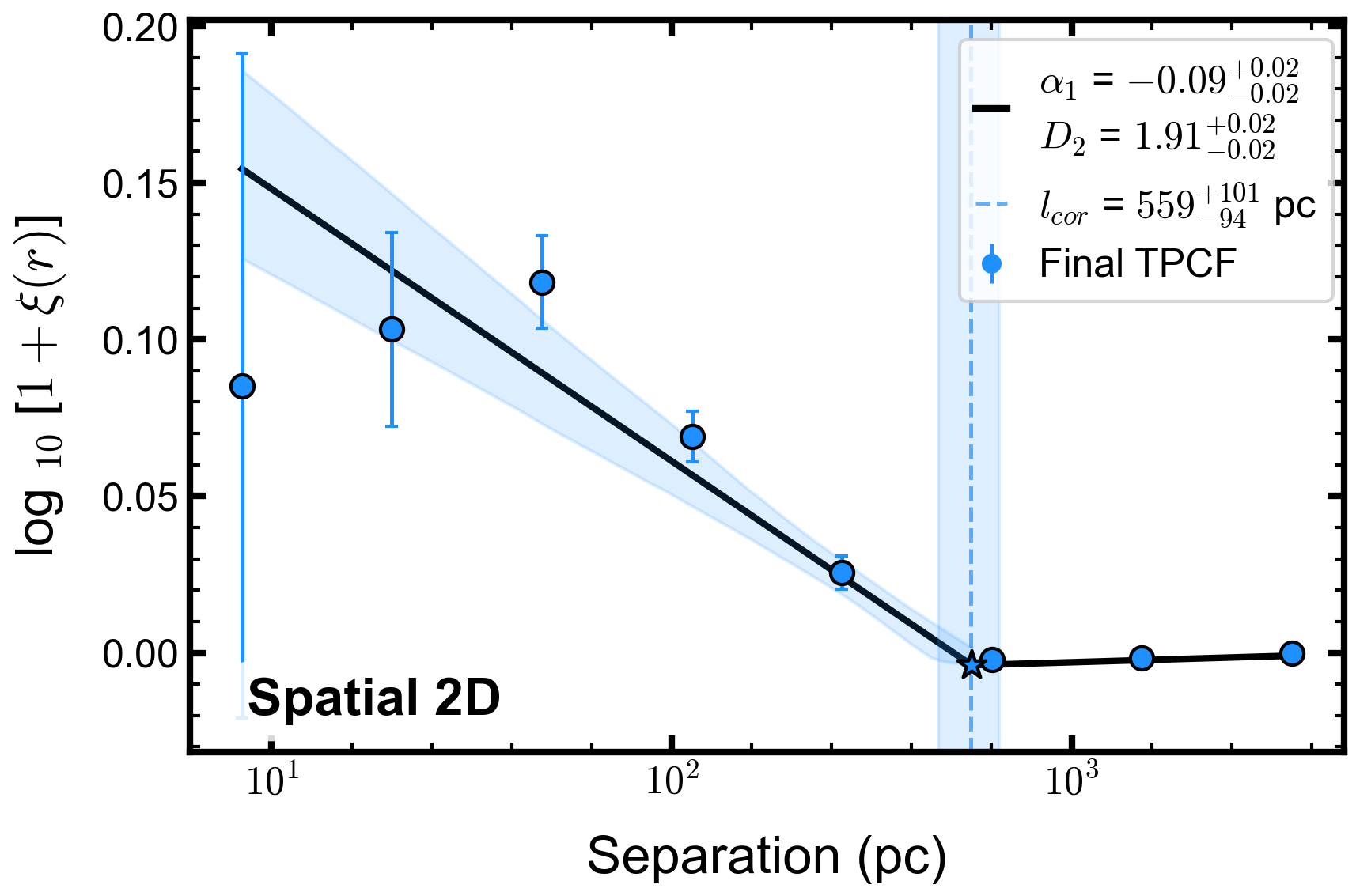}
    \includegraphics[width=0.48\linewidth]{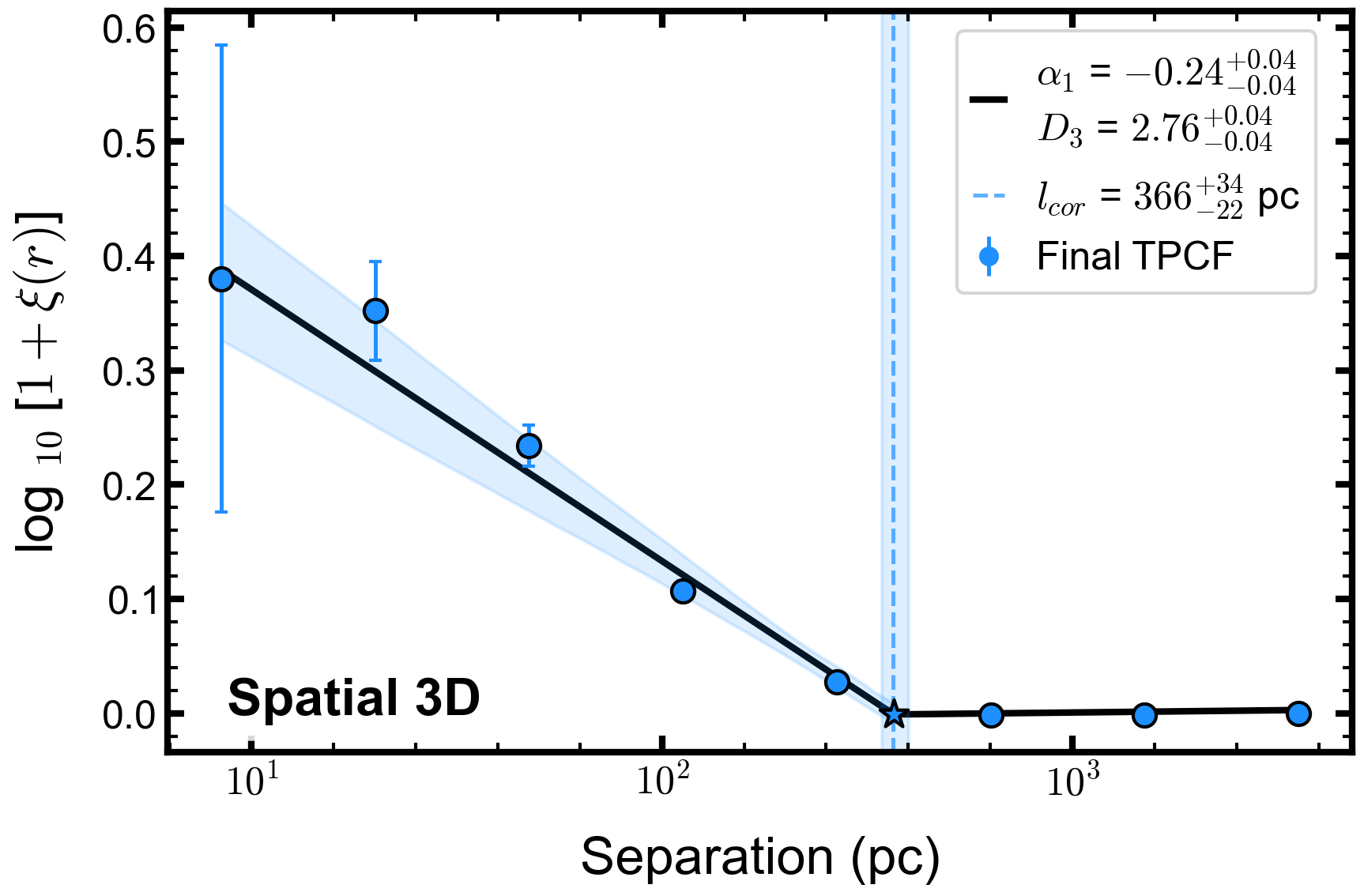}
    \caption{TPCFs for all OC samples in in 2D (left) and 3D (right) sapce. Circles mark the median values adopted for the fitting. Dashed lines indicate the breakpoints of the fitted TPCF.}
    \label{fig: overall_space}
\end{figure*}

Both the 2D and 3D analyses yield strongly clustered distributions, as shown in the TPCF results in \figref{fig: overall_space}. After converting the $D_3$ to a common $D_2$ scale using \equref{equ: cal}, the inferred $D_2$ is 1.91, in agreement with the directly measured 2D value $D_2$=1.91. This close agreement indicates that projection effects have only a minor influence on the global clustering properties of nearby OCs.
The median values of $l_{\rm cor,3D}$ are about 370 pc in 3D and $l_{\rm cor,2D} \sim$560~pc in 2D. The larger value obtained in projection suggests that the vertical structure of the Galactic disk primarily affects the characteristic clustering scale rather than the inferred clustering strength.

%-------------------------- velocity
\subsection{Velocity clustering in the solar neighborhood}
\begin{figure*}
    \centering    
    \includegraphics[width=0.48\linewidth]{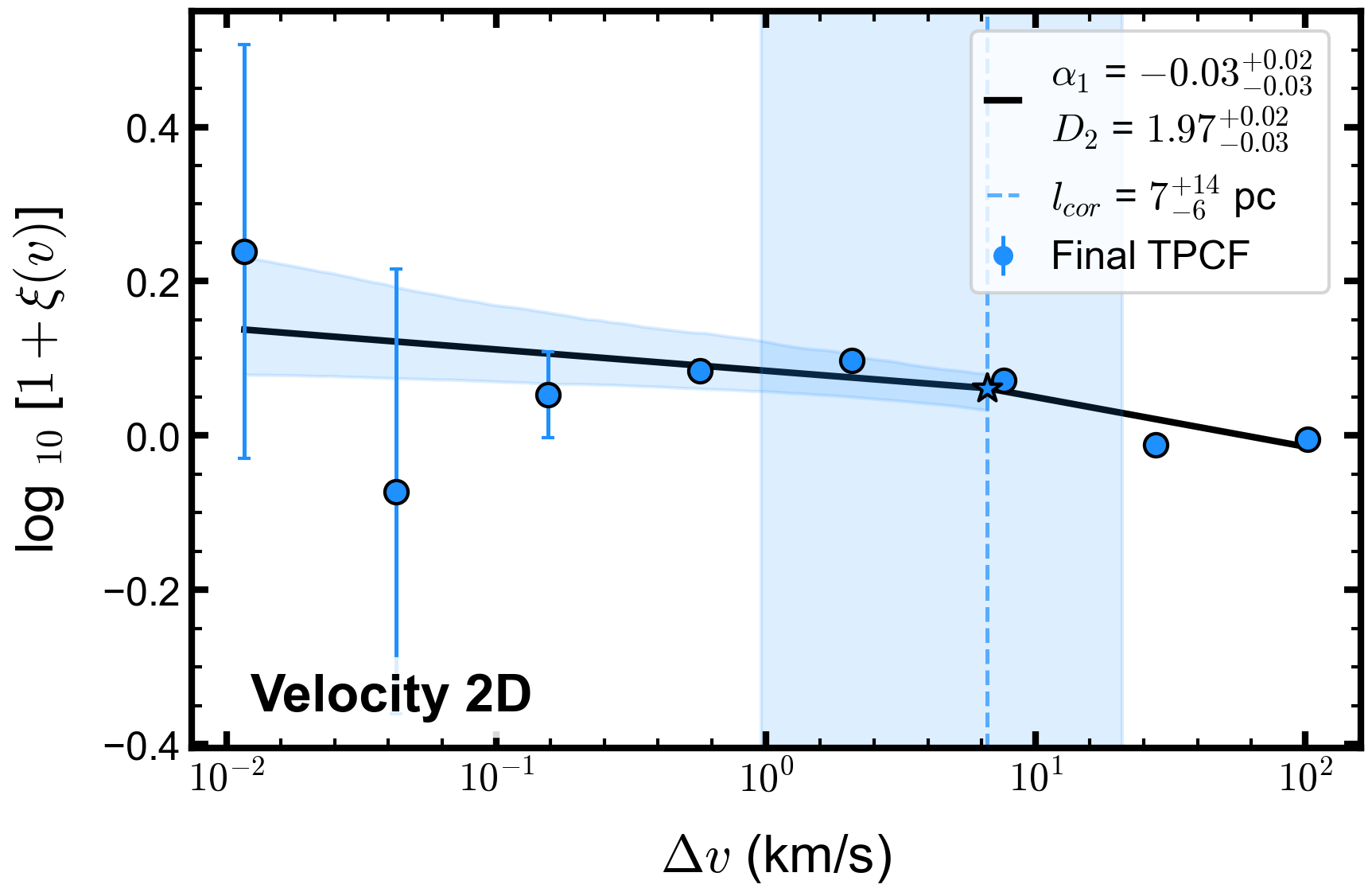}
    \includegraphics[width=0.48\linewidth]{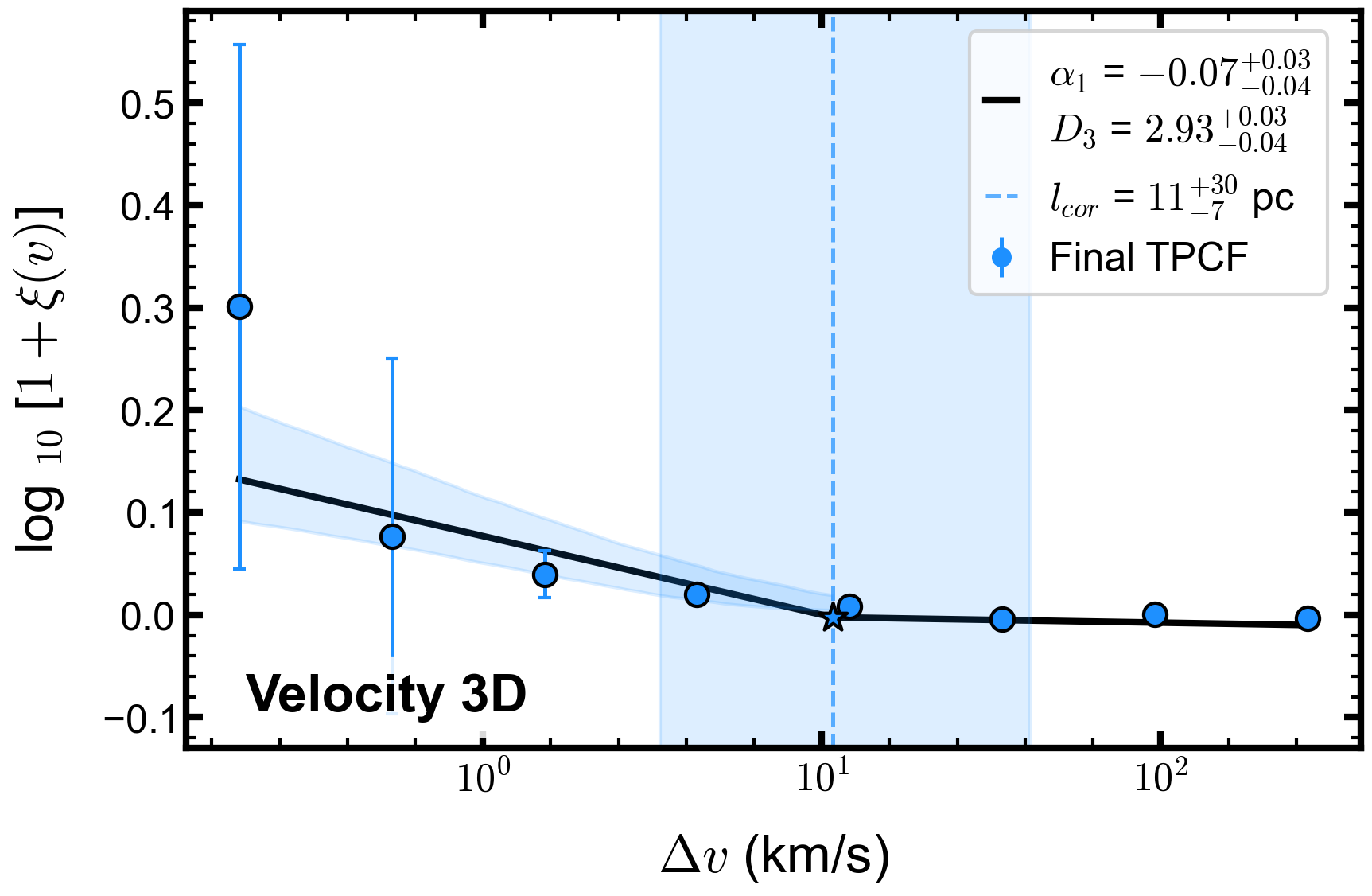}
    \caption{TPCFs for all OC samples in 2D (left) and 3D (right) velocity. Circles mark the median values adopted for the fitting. Dashed lines indicate the breakpoints of the fitted TPCF.}
    \label{fig:overall_velocity}
\end{figure*}

To characterize the kinematic clustering of OCs, we computed the TPCF in velocity space using both the 3D Galactocentric velocity vectors ($U,V,W$) and the tangential velocity ($V_t$).
Both the 2D and 3D velocity TPCFs exhibit a statistically significant excess at small velocity separations (as shown in \figref{fig:overall_velocity}), indicating the presence of kinematic clustering. However, the inferred fractal dimensions ($D_3=2.93$ and $D_2$=1.97 after 3$\sigma$ clipping) are substantially closer to the values expected for a random distribution than those measured in spatial coordinates, implying that the velocity hierarchy is considerably weaker than the spatial hierarchy. The characteristic velocity correlation scales are $l_{\rm cor}\approx$11 \kms\ and 7 \kms\ for the 3D and 2D analyzes, respectively. These scales represent the characteristic extent of coherent structures in velocity
space, below which groups of clusters remain kinematically correlated. Beyond this scale, the velocity distribution becomes statistically indistinguishable from a random population.

\begin{table}[]
\centering
\renewcommand{\arraystretch}{1.2}
\caption{2D and 3D TPCF results for different age bins}
\label{tab:age_results}
\begin{tabular}{lrcccc}
\toprule
Age & $N$ & $D_2$ & $l_{\rm cor,2D}$ & $D_3$ & $l_{\rm cor,3D}$ \\
(Myr) & & & (pc)&&(pc)\\
\midrule
$<$9 & 168 & $1.45_{-0.21}^{+0.14}$ & $363_{-84}^{+116}$ & $2.02_{-0.42}^{+0.27}$ & $262_{-74}^{+76}$ \\
9--30 & 319 & $1.79_{-0.03}^{+0.02}$ & $648_{-96}^{+57}$ & $2.61_{-0.12}^{+0.10}$ & $471_{-104}^{+302}$ \\
30--96 & 636 & $1.92_{-0.06}^{+0.05}$ & $757_{-349}^{+528}$ & $2.76_{-0.09}^{+0.07}$ & $532_{-156}^{+358}$ \\
96--309 & 1037 & $2.05_{-0.12}^{+0.06}$ & $475_{-196}^{+467}$ & $2.95_{-0.03}^{+0.01}$ & $616_{-177}^{+99}$ \\
309--991 & 403 & $2.31_{-0.05}^{+0.05}$ & $129_{-9}^{+11}$ & $2.44_{-0.22}^{+0.19}$ & $70_{-22}^{+36}$ \\
991--6363 & 83 &    --       &      --     &          -- & --         \\
\bottomrule
\end{tabular}
\end{table}

\begin{table*}
\centering
\caption{Structural parameters derived from the TPCF for different mass and distance bins.}
\label{tab:mass_dist_results}
\renewcommand{\arraystretch}{1.2}
\begin{tabular}{lrrccrcc}
\toprule
Range &$N_{\leq100}$ &$D_{3,\leq100}$ &$l_{{\rm cor},\leq100}$ &$N_{\rm all}$ &$D_{3,\rm all}$ &$l_{{\rm cor},\rm all}$ \\
 &&&(pc) &&&(pc) \\
\midrule
\multicolumn{7}{l}{{Mass range ($M_{\odot}$)}}\\
\midrule
40--136 & 84 & $1.97_{-0.44}^{+0.39}$ & $136_{-22}^{+55}$ & 159 & $2.37_{-0.25}^{+0.23}$ & $164_{-29}^{+127}$ \\
136--460 & 227 & $2.55_{-0.13}^{+0.13}$ & $401_{-62}^{+237}$ & 469 & $2.76_{-0.09}^{+0.05}$ & $586_{-217}^{+272}$ \\
460--1557 & 181 & $2.70_{-0.07}^{+0.06}$ & $967_{-217}^{+265}$ & 424 & $2.82_{-0.02}^{+0.02}$ & $865_{-65}^{+173}$ \\
1557--5271 & 59 & $2.46_{-0.17}^{+0.15}$ & $305_{-72}^{+85}$ & 184 & $2.75_{-0.07}^{+0.06}$ & $367_{-85}^{+152}$ \\
5271--17849 & 11 & $2.39_{-0.44}^{+0.30}$ & $688_{-270}^{+444}$ & 30 & $2.54_{-0.19}^{+0.15}$ & $542_{-190}^{+528}$ \\
\midrule
\multicolumn{7}{l}{{Distance range (pc)}}\\
\midrule
47--1637 & 305 & $2.66_{-0.18}^{+0.10}$ & $576_{-390}^{+493}$ & 715 & $2.84_{-0.03}^{+0.03}$ & $1063_{-108}^{+205}$ \\
1637--2063 & 118 & $2.27_{-0.20}^{+0.36}$ & $815_{-230}^{+297}$ & 245 & $2.47_{-0.14}^{+0.20}$ & $917_{-199}^{+346}$ \\
2063--2362 & 80 & $2.41_{-0.16}^{+0.14}$ & $1003_{-258}^{+243}$ & 163 & $2.46_{-0.15}^{+0.14}$ & $1064_{-211}^{+259}$ \\
2362--2599 & 44 & $2.78_{-0.23}^{+0.23}$ & $588_{-377}^{+564}$ & 92 & $2.54_{-0.15}^{+0.27}$ & $1003_{-782}^{+373}$ \\
2599--2800 & 20 & $2.74_{-0.17}^{+0.30}$ & $497_{-333}^{+574}$ & 48 & $2.48_{-0.10}^{+0.27}$ & $824_{-254}^{+371}$ \\
\bottomrule
\end{tabular}
\end{table*}

%===========================================
% discussion
%===========================================
\section{Discussion} \label{sec:discussion}
\subsection{The hierarchical imprint of star formation}
The youngest OCs in our sample ($<$9 Myr) exhibit the strongest clustering signal, with a 2D fractal dimension of $D_2 \approx 1.45^{+0.14}_{-0.21}$, corresponding to $D_3 \approx 2.02^{+0.27}_{-0.42}$ in 3D. Since values of
$D_2<2$ and $D_3<3$ indicate clustering (\secref{sec:method_fractalD}), these values indicate a highly hierarchical spatial distribution.

Our results are broadly consistent with the study of \citet{Alfaro2022ApJ...937..114A}, who analyzed 309 young OCs within 3.5 kpc of the Sun and measured a 2D fractal dimension of $D_2 \approx 1.62$ for systems younger than $\sim31$ Myr. The slightly lower fractal dimensions found here likely reflect our ability to isolate the youngest evolutionary stages more effectively. The larger and more complete catalog \citetalias{Hunt2024} allows a finer age division while maintaining robust statistics, providing a more direct view of the primordial clustering regime and reducing contamination from dynamically evolved systems.

The fractal dimensions measured for the youngest clusters are remarkably similar to those reported for the ISM (e.g. $D_2 \sim$ 1.2--1.6, \citealt{Beech1987, Falgarone1991, Elmegreen1996, Sanchez2005, Elmegreen2006, Sanchez2008}).
This agreement supports the spatial inheritance scenario of star formation, in which newly formed clusters retain the hierarchical structure of giant molecular clouds \citep{Elmegreen_Salzer1999AJ....117..764E,Elmegreen_Scalo2004ARA&A..42..211E,Elmegreen2003ApJ...593..333E,Elmegreen2007ApJ...668.1064E,Klessen2009ApJ...692..364F}. Such a connection is naturally expected if star formation proceeds within a turbulent and self-similar interstellar medium, where stellar structures initially inherit their fractal morphology.

\subsection{Characteristic fading timescale of hierarchical structure}
Although the youngest OCs preserve a clear imprint of the hierarchical structure of their natal gas, our TPCF analysis demonstrates that the clustering signature weakens systematically with time. Both the 2D and 3D measurements reveal a transition at a characteristic age of approximately 100 Myr, beyond which the clustering signal becomes substantially weaker and the spatial distribution approaches a nearly random configuration. This result suggests that the hierarchical structure remains detectable for only a finite period before being progressively dissolved \citep{Grasha2015ApJ...815...93G,Grasha2017_5,Elmegreen_2018ApJ...853...88E,Grasha2019MNRAS}. Such behaviour is expected as
initially correlated stellar structures gradually lose their spatial coherence owing to dynamical evolution \citep{Elmegreen_2018ApJ...853...88E}.

The measured age--scale interaction shows that the spatial hierarchy is erased progressively from small to large scales. As cluster age increases, the excess correlation at separated by tens of parsecs decreases more rapidly than the excess at separations of hundreds of parsecs. The TPCF therefore becomes flatter: compact clustering fades first, while weaker correlations on larger scales remain detectable for longer. This is a statement about the relative evolution of the clustering signal. It does not imply that clustering power is conserved or physically transferred from small to large scales, because the analysis does not follow the same cluster pairs through time.

The inferred dissolution timescale is broadly consistent with  measurements based on star clusters in nearby galaxies, which typically report values of $\sim$40-100 Myr \citep{Grasha2017_5,Elmegreen_2018ApJ...853...88E,Grasha2019MNRAS}. Similar values have also been found in other galactic environments, indicating that the gradual dispersal of hierarchical stellar structures is a common outcome of cluster evolution.
At the same time, the diversity of dissipation timescales reported among different galaxies supports the idea that the erasure of hierarchical structure is influenced by the galactic environment \citep{Menon2021MNRAS.507.5542M}.

\subsection{Weak mass and volume dependence}
In contrast to the strong age dependence discussed above, the clustering properties of OCs exhibit only weak variations with cluster mass and heliocentric distance. After accounting for the mass--distance completeness relation from \citetalias{Hunt2024}, neither the fractal dimension nor the correlation length shows a statistically significant monotonic dependence on mass or survey volume.

The weak mass dependence is consistent with the extragalactic studies \citep{Grasha2015ApJ...815...93G,Grasha2017_5}, which found that cluster mass has only a minor direct influence on clustering strength. Within the framework of turbulence-driven hierarchical star formation, the spatial distribution of stellar structures is expected to be self-similar across a wide range of masses \citep{ElmegreenFalgarone1996,Mac-Low2004,Grasha2015ApJ...815...93G}.
In our sample, the lowest-mass bin exhibits a marginal enhancement in clustering strength relative to the higher-mass bins. Nevertheless, the significance of this excess is modest compared to the strong age dependence observed throughout the sample and remains comparable to the measurement uncertainties. We therefore conclude that the current data provide no robust evidence for an intrinsic mass dependence of the clustering hierarchy.

Likewise, the absence of a systematic dependence on heliocentric distance demonstrates that the age evolution reported in this study is unlikely to be an artifact of residual observational selection effects. The similarity of the clustering measurements across the surveyed volume further suggests that the hierarchical structure of OCs is broadly uniform throughout the solar neighborhood.

Taken together, these results indicate that age is the dominant parameter governing the observed evolution of clustering strength. Neither cluster mass nor survey volume produces variations comparable to those associated with the progressive dynamical dispersal of hierarchical structure \citep{Krumholz2019ARA&A..57..227K}.

\begin{figure*}
    \centering
    \includegraphics[width=0.48\linewidth]{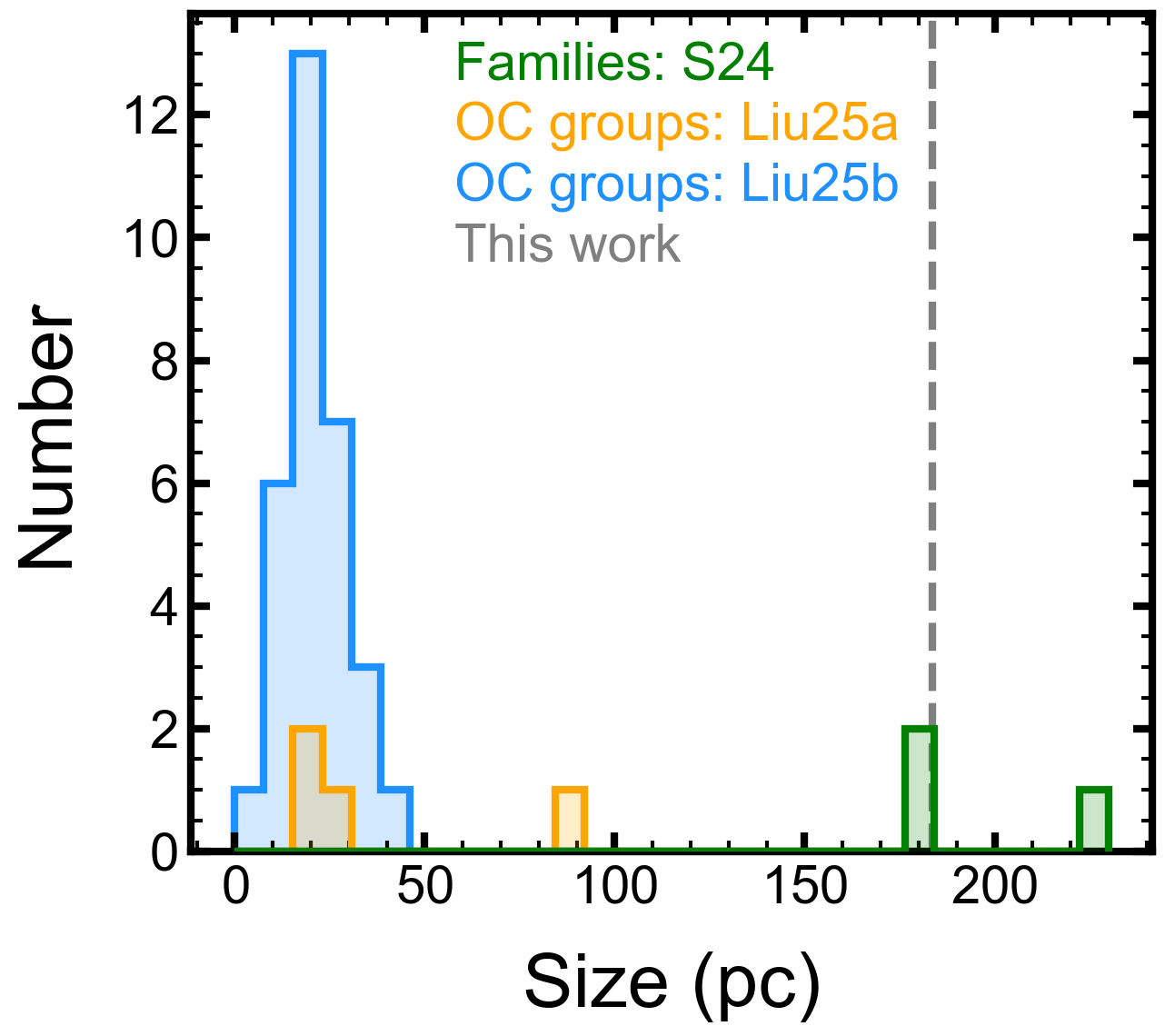}
    \includegraphics[width=0.48\linewidth]{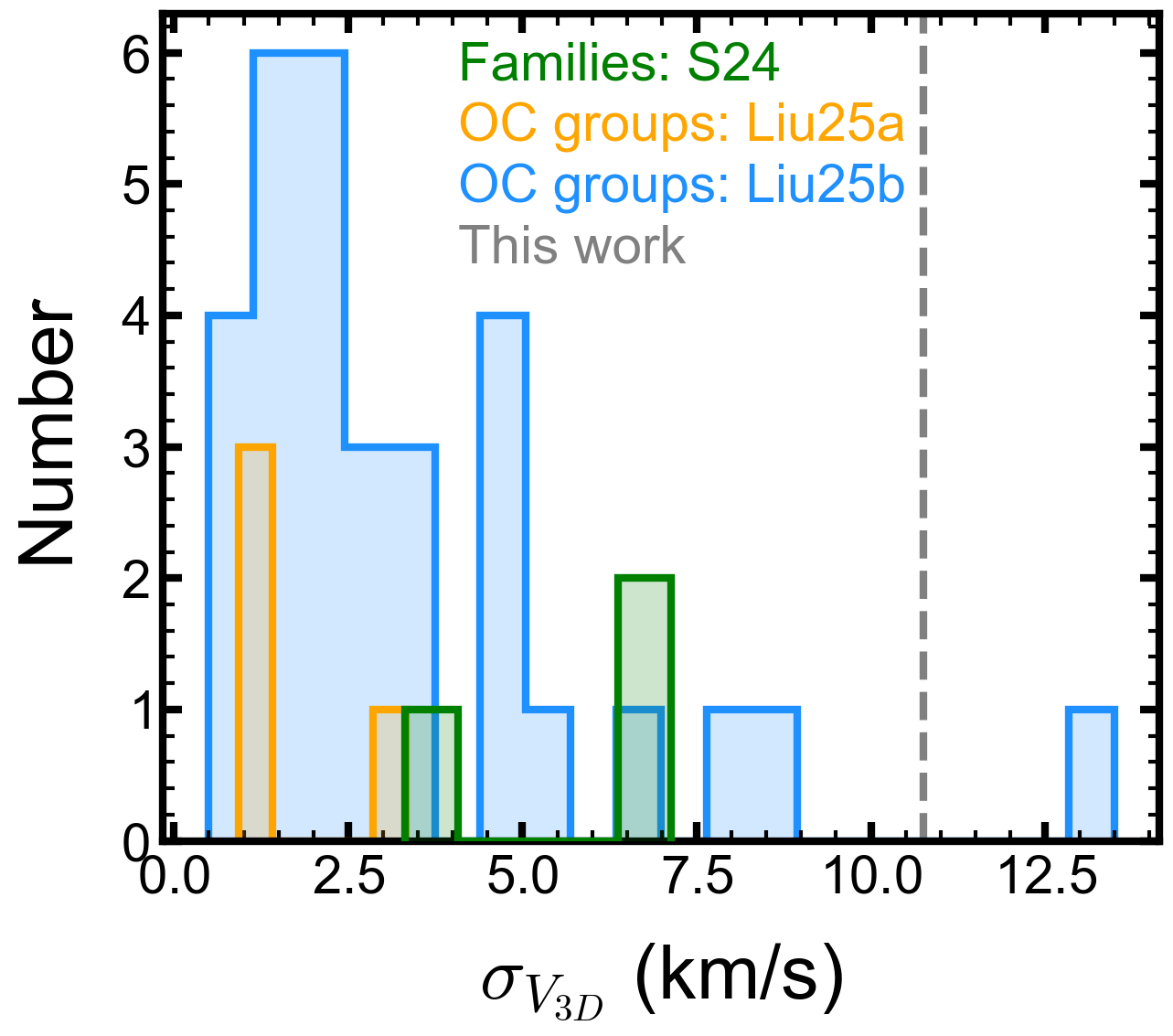}
    \caption{Distribution of physical sizes and velocity dispersions for OC families/groups. The left panel illustrates the distribution of physical sizes for cluster families and groups identified in recent literature, including families from \citet{Swiggum2024Natur.631...49S} shown in green and OC groups from \citealt{Liu2025a, Liu2025b} shown in orange and blue. The right panel displays the velocity dispersion ($\sigma_{V_{3D}}$, in \kms) for these OC groups and binary clusters. The grey dashed line marks half of the $l_{cor}$ value derived in this work.
    }
    \label{fig: size_vsigma}
\end{figure*}

\subsection{Empirical calibration between $D_2$ and $D_3$}
Because observations of external galaxies are generally limited to projected coordinates, establishing a connection between intrinsic 3D ($X,Y,Z$) fractal dimensions and 2D ($X,Y$) projected is essential for interpreting clustering measurements in a common framework. 
We utilized our measurements of self-similar OC distributions in 2D and 3D to derive an empirical relation between $D_3$ and $D_2$ in the Milky Way (as shown in \figref{fig: ageD_cali}). 

Our results demonstrate that the relation between $D_2$ and $D_3$ is not simply given by $D_3 = D_2 + 1$, but is instead more complex, as has often been suggested \citep{Gouliermis2014MNRAS_ngc346, Grasha2015ApJ...815...93G}.
This is particularly important because previous studies rely on 2D data from extragalactic star clusters, for which the relation to the intrinsic 3D structure is inferred from simulations \citep{Gouliermis2014MNRAS_ngc346}. The Milky Way offers a unique opportunity to measure both intrinsic 3D fractal dimensions and 2D projected dimensions.
Our observationally calibrated relation based on direct 3D measurements in the Milky Way  therefore provides an observational benchmark for connecting 2D and 3D fractal dimensions and complements existing simulation-based approaches.

We emphasize that the relation derived here should be regarded as an empirical calibration for the OCs in this work rather than a universal mapping between $D_2$ and $D_3$. Nevertheless, it provides a physically motivated framework for placing intrinsic 3D fractal dimensions and 2D projected clustering measurements on a common scale and enables more direct comparisons between studies based on different observational geometries.

\subsection{Characteristic scales of intermediate Galactic structures}

We find that OCs within 2.8 kpc exhibit clustering in both spatial and velocity components. The characteristic correlation lengths in 3D spatial and velocity components are $\sim$370~pc and $\sim$11~\kms, respectively. As illustrated in \figref{fig: size_vsigma}, these characteristic scales are comparable to the physical sizes and velocity dispersions of recently identified cluster families and OC groups \citep{Swiggum2024Natur.631...49S,Liu2025a, Liu2025b}, providing an independent support that the TPCF traces the same physical structures.
The comparison requires accounting for the different definitions of size. The sizes of cluster families and OC groups are characterized by their radii, defined as the median distance of member clusters from the group center, whereas the correlation length measures the maximum scale over which clustering remains significant and is therefore more directly comparable to the family diameter. The reported family radii of $\sim170$--230 pc correspond to diameters of $\sim340-460$ pc, in excellent agreement with the measured $l_{\rm cor,3D}\simeq370$ pc. 

The interpretation of $l_{\rm cor}$ as the characteristic size of cluster families is further supported by the orbital traceback analysis presented in \appref{sec:appendix}. We recomputed the TPCF at each integration timestep and measured the corresponding correlation length. For both cluster families, $l_{\rm cor}$ increases progressively with time, consistent with the gradual expansion of the families reported by \citet{Swiggum2024Natur.631...49S}.
In contrast, the fractal dimension $D_3$ remains nearly unchanged throughout the orbital evolution, indicating that dynamical evolution primarily modifies the physical size of the clustered structures rather than their intrinsic hierarchical morphology. The agreement between the present-day correlation length and the evolution tendency of the size of the families over time therefore suggests that $l_{\rm cor,3D}$ traces a physically meaningful characteristic scale of coherent star-forming structures.

The spatial correlation scale derived here is also consistent with other tracers of Galactic stellar clustering.
\citet{Hinkel2023ApJ...942...41H} identified sub-kiloparsec spatial clustering among field stars within a 3~kpc region, while \citet{Kamdar2021ApJ} reported a characteristic clustering scale of $\sim300$ pc. The consistency between these two independent tracers strongly suggests that star formation within the spiral arms is not a collection of isolated local events but rather proceeds in a hierarchical manner within large-scale structures extending over hundreds of parsecs to kiloparsec scales.

However, the characteristic velocity scale inferred from OCs ($\sim11$ \kms) is smaller than the $\sim15$ \kms measured for field stars by \citet{Kamdar2021ApJ}.
This discrepancy likely reflects the distinct nature of the tracers. OCs represent relatively young and dynamically coherent systems that remain kinematically coeval. As a result, their internal and relative velocities remain low, leading to a smaller characteristic clustering scale in velocity space. In contrast, field stars consist of a mixture of populations spanning a wide range of ages and dynamical histories. Consequently, OCs provide a cleaner and more direct probe of the large-scale star-forming hierarchy from which they originated. We note, however, that RV uncertainties are not explicitly accounted for, and the RVs of OCs with only a limited number of member stars with RV measurements may be less reliable. Such uncertainties may broaden the observed velocity-space distribution and affect the inferred characteristic velocity scale. The improved precision and increased number of stellar RV measurements expected from \gdr{4} will allow this uncertainty to be better constrained in future analyses.

We find a notable difference in $l_\text{cor}$ between the 2D and 3D analyses. The 2D TPCF yields a characteristic scale of $l_{\rm cor,2D}\approx$560~pc, substantially greater than $l_{\rm cor,3D}$ ($\approx$370~pc). 
The $l_{\rm cor,2D}$ derived from the 2D analysis is broadly consistent with values reported for nearby galaxies ($\sim$80--500 pc, \citealt{Menon2021MNRAS.507.5542M,Grasha2017_5}). This scale may be associated with the transition between 2D projected and intrinsic 3D structures, as well as the finite vertical extent of the Galactic disk, although its physical interpretation remains unclear \citep{Gouliermis2014MNRAS_ngc346,Grasha2015ApJ...815...93G}. A major limitation of these studies is that only projected spatial information is available, making it difficult to distinguish intrinsic 3D structure from projection along the $Z$ direction. The Milky Way offers a unique opportunity to address this issue because full 3D positions are available for OCs. 

The 2D analysis yields $l_{\rm cor,2D}$, which is comparable to the vertical extent of the Galactic disk, lying between the canonical scale heights of the thin ($\sim$300 pc) and thick ($\sim$900 pc) disk components \citep{Juric2008ApJ,BlandHawthorn2016ARA&A..54..529B}. This suggests that the break scale measured in 2D TPCF may influenced not only by the intrinsic clustering hierarchy but also by the finite thickness of the disk along the vertical extent of the Galactic disk..
In contrast, the $l_{\rm cor,3D}$ is less affected by projection effects and likely represents the maximum physical scale over which OCs remain associated within coherent star-forming structures in the solar neighborhood.

%===========================================
% summary
%===========================================
\section{Summary} \label{sec:summary}
In this work, we investigate the first systematic characterization of the 3D hierarchical structure of open clusters across the solar neighborhood using the homogeneous catalog of \citetalias{Hunt2024}.
By applying the TPCF to high-quality OC samples, and adopting the mass-distance completeness framework from \citetalias{Hunt2024}, we characterize the hierarchical spatial and kinematic structure of  OCs, quantify its evolution with time, and investigate its connection to large-scale star-forming structures in the Milky Way.

Our main conclusions are as follows.
The clustering strength of OCs declines systematically with increasing age. Combining the 2D and 3D TPCF measurements, we find that the hierarchical clustering signal becomes weak on a characteristic timescale of order 100 Myr. This timescale quantifies the progressive loss of spatial correlations among Galactic open clusters. 
The present age-binned analysis constrains the approximate fading timescale but does not establish a sharp transition at exactly 100 Myr.
The fractal dimension $D_2$ measured for the youngest clusters is comparable to that observed in the interstellar medium, indicating that the spatial distribution of young clusters retains the primordial imprint of the hierarchical fragmentation of their natal molecular clouds and thus provides a direct link between ISM and stellar clustering.

In contrast to the strong age dependence, the clustering strength shows only weak variations with cluster mass and heliocentric distance, of which no monotonic trend is detected across the mass and distance ranges in this study. These results indicate that age is the dominant parameter governing the evolution of clustering strength, while cluster mass and heliocentric distance play only secondary roles.
By comparing the 2D and 3D clustering strength measurements, we establish an empirical relation between the projected and intrinsic fractal dimension for Milky Way, $D_2 \sim 0.60 \times D_3+0.25$. This calibration provides a useful bridge between Galactic studies based on full 3D information and extragalactic studies that rely on projected cluster distributions.
 
OCs exhibit a typical spatial clustering scale of approximately 370~pc in 3D and a characteristic velocity scale of 11~\kms. These scales are consistent with the sizes and velocity dispersions of recently identified cluster families, providing an independent statistical characterization of the large-scale hierarchy of star formation in the Milky Way.

Overall, these results provide a coherent observational picture of hierarchical star formation and subsequent dynamical evolution in the Galactic disk. OCs emerge as effective tracers of both the primordial fractal structure of the ISM and the long-term dynamical mixing processes operating within the Milky Way.

%------------------
%acknowledgements
%------------------
\begin{acknowledgements}
% % We are grateful to an anonymous referee for valuable comments, which have improved the paper significantly. 
JA and EM acknowledges co-funding by the European Union (ERC, ISM-FLOW, 101055318). Views and opinions expressed are, however, those of the author(s) only and do not necessarily reflect those of the European Union or the European Research Council. Neither the European Union nor the granting authority can be held responsible for them.
YZ acknowledges the support from the science research grants from the Chinese Academy of Sciences (CAS) "Light of West China" Program (No. 2022-XBQNXZ-013) and Central Guidance for Local Science and Technology Development Fund (No. ZYYD2025QY27).
JG acknowledges co-funding from the European Union, the Central Bohemian Region, and the Czech Academy of Sciences, as part of the MERIT fellowship, MSCA-COFUND Horizon Europe, Grant agreement 101081195. Views and opinions expressed are, however, those of the author(s) only and do not necessarily reflect those of the European Union or the European Research Council. Neither the European Union nor the granting authority can be held responsible for them.
GL would like to thank the China Scholarship Council (CSC) for support (Grant Nos. 202504910181)
This work has made use of data from the European Space Agency (ESA) mission \gaia \url{https://www.cosmos.esa.int/ gaia}, processed by the \gaia Data Proces-Processingalysis Consortium (DPAC,\url{https://www.cosmos. esa.int/web/gaia/dpac/consortium}). Funding for the DPAC has been provided by national institutions, in particular, the institutions participating in the \gaia Multilateral Agreement.
\end{acknowledgements}

\bibliographystyle{aa}
\bibliography{reference}
%--------------------------------------------------------------------------------
\begin{appendix}
\section{Evolution of clustering during orbital traceback}\label{sec:appendix}
\subsection{Star Cluster Families}\label{sed: data_family}
To distinguish between global clustering behavior and the intrinsic dynamical evolution of physically related systems, we incorporate the three massive families in the solar neighborhood identified by \citealt{Swiggum2024Natur.631...49S} (hereafter \citetalias{Swiggum2024Natur.631...49S}). These families originated from three compact and massive star-forming complexes, and their large numbers of member clusters provide a statistically robust sample for studying the evolution of clustering properties and the preservation of hierarchical structure over time. The three families are referred to as the Collinder 135 (Cr135), Messier 6 (M6), and Alpha Persei ($\alpha$\,Per) families.
\begin{itemize}
    \item M6 family: 34 member OCs, with the youngest cluster having an age of 14.6 Myr.
    \item Cr135 family: 39 member OCs, with the youngest cluster having an age of 6.8 Myr.
    \item $\alpha$\,Per family: 82 member OCs, making it the largest family in the sample. However, its star formation process was completed only $\sim1$ Myr ago.
\end{itemize}

\subsection{Orbital Traceback}
To investigate the temporal evolution of the clustering, we integrated their Galactic orbits backward in time and measured the TPCF over time. 
The orbit integration was performed using the \texttt{galpy} Python package \citep{Bovy-2015ApJS} with the axisymmetric Galactic potential model \texttt{MWPotential2014}, which includes disk, bulge, and halo components.
For the 2289 OCs described in \secref{data:OCs}, we further restrict the sample to clusters younger than 100 Myr, yielding a final sample of 1167 OCs. The initial conditions consist of the 6D phase-space coordinates (\RAdeg, \DEdeg, \plx, \pmra, \pmdec, RV) and age from \citetalias{Hunt2024}.
For the families, we adopt the heliocentric positions and velocities published by \citetalias{Swiggum2024Natur.631...49S}.
Because the extremely young age of the $\alpha$\,Per family limits the significance of orbital traceback, it is excluded from the subsequent dynamical analysis. The remaining dynamical investigation therefore focuses on the combined sample of 73 clusters from the M6 and Cr135 families.

To ensure that clustering measurements at different epochs are directly comparable, all samples are traced back only to the formation epoch of the youngest member in each sample. Extending the integration further back would progressively remove younger clusters from the sample, leading to a time-dependent membership and consequently altering the TPCF simply because of changing sample composition. Restricting the traceback interval in this way ensures that the same cluster population is used at all epochs, allowing the evolution of the clustering properties to be measured consistently. A consequence of this approach is that the maximum traceback time is limited by the age of the youngest cluster in each sample. In particular, the youngest cluster in the young sample has an age of only 2.9 Myr, restricting the traceback analysis of the full young-cluster sample to the most recent $\sim$3 Myr of evolution.

To quantify the spatial evolution of each sample during the orbital traceback, we measured the projected ellipticity in the Galactic $X$--$Y$ plane at each timestep. We characterized the projected morphology using the covariance ellipse derived from each moment of the spatial distribution, a method widely adopted in studies of stellar systems and galaxy morphology \citep{Carter1980MNRAS.191..325C}.
The sample center was defined by the median position of all clusters, and the principal axes were obtained from the eigenvalues of the covariance matrix. The projected ellipticity ($e$) was then calculated as
$e=1-\frac{b}{a},$
where $a$ and $b$ are the lengths of the semi-major and semi-minor axes, respectively. By definition, $e=0$ corresponds to a circular distribution, while larger values indicate increasingly elongated morphologies. This procedure was repeated at every traceback timestep to quantify the evolution of the projected morphology

% %-------------------------------------------------
\subsection{Tracing the dynamical evolution of clustering}
\begin{figure*}
    \centering
    \includegraphics[width=0.99\linewidth]{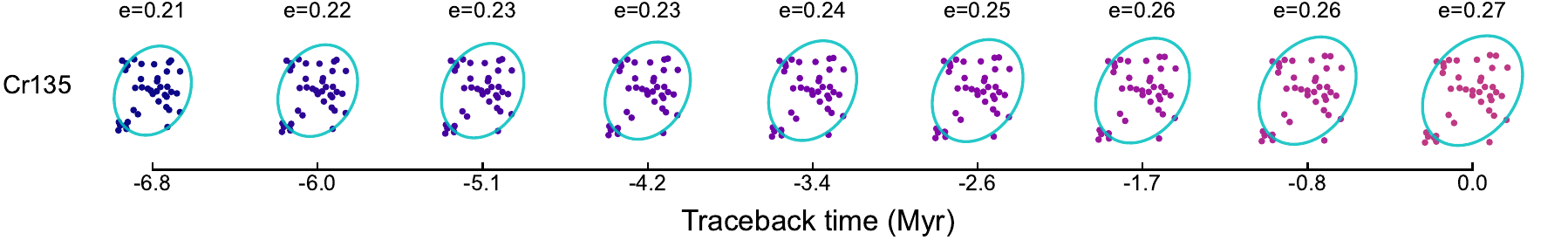}
    \includegraphics[width=0.99\linewidth]{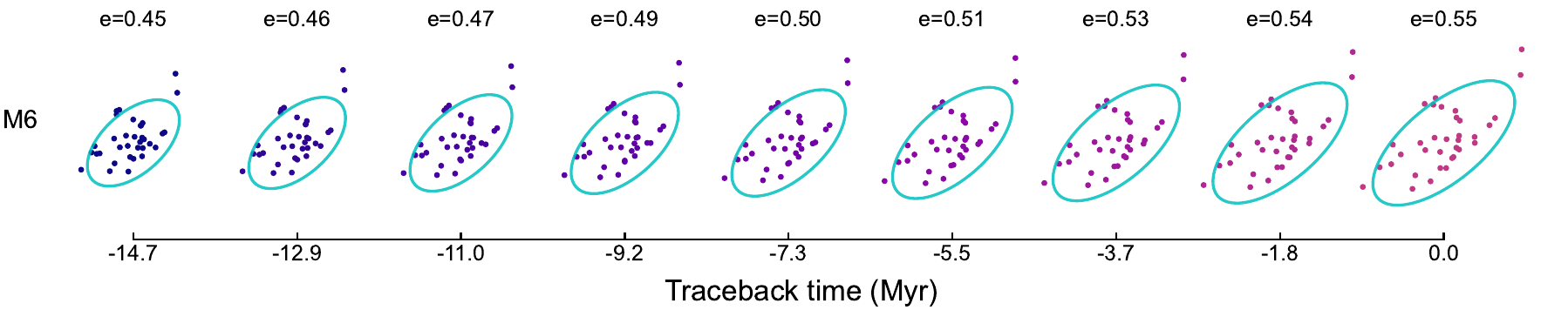}
    \caption{Spatial-temporal evolution of cluster families Cr135 and M6. Families become more extended and elongated over time (see also \citetalias{Swiggum2024Natur.631...49S})} 
    % piexl:2400*460
    \label{fig: track_visual}
\end{figure*}

\begin{figure*}
    \centering
\includegraphics[width=0.95\linewidth]{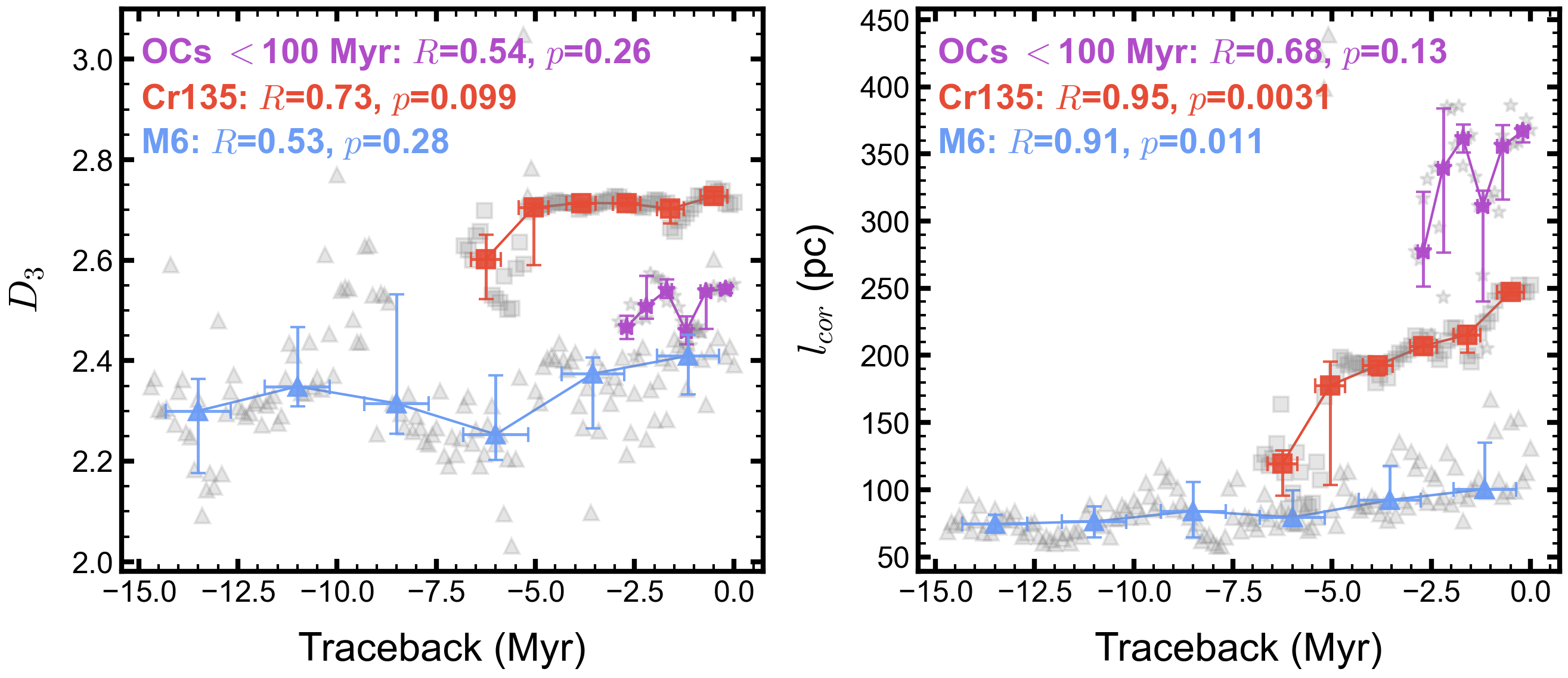}
    \caption{The evolution of the fractal dimension $D_3$ and the correlation length $l_{\rm cor,3D}$ as a function of traceback time once all clusters are formed. Grey points show the $D_3$ values measured at each individual time step, while colored symbols indicate the values in each bin. Purple symbols show the results for the entire sample, while red and blue symbols correspond to the Cr135 and M6 families, respectively. Error bars represent the 16th and 84th percentiles within each bin, corresponding to the 1$\sigma$ confidence interval. }
    \label{fig: traceback_D3}
\end{figure*}

Snapshots of the orbital traceback for the young OCs and two OC families (Cr135 and M6) at different epochs are shown in \figref{fig: track_visual}, together with the ellipse fitting at each time step. The two families become progressively more elongated over time. 
In all families samples, the systems occupy progressively more compact configurations at earlier epochs, while their overall hierarchical morphology remains largely preserved. This behavior is quantified in \figref{fig: traceback_D3}, where the evolution of $D_3$ and $l_{\rm cor}$ is presented as a function of traceback time.

Across all samples, the evolution of $D_3$ is considerably weaker than that of $l_{\rm cor}$. The fractal dimension remains approximately constant within the uncertainties, indicating that the degree of hierarchical clustering changes little over the traceback interval. In contrast, $l_{\rm cor}$ decreases systematically toward earlier times, demonstrating that the cluster distributions were substantially more compact in the past and have expanded over time. This trend is consistent with the expansion of cluster families reported by \citetalias{Swiggum2024Natur.631...49S}. By comparison, $D_3$ remains nearly unchanged throughout the orbital evolution, indicating that dynamical evolution primarily modifies the physical scale of clustered structures rather than their intrinsic hierarchical morphology.

The strongest evolution is observed for the two cluster families, for which statistically significant correlations between $l_{\rm cor}$ and traceback time are detected (\figref{fig: traceback_D3}).  The young cluster sample also exhibits a larger $l_{\rm cor}$ than individual families, reflecting its greater spatial extent as a mixture of multiple dynamically distinct structures.
Across all samples, the evolution of $D_3$ is less pronounced than that of $l_{\rm cor}$, as shown in \figref{fig: traceback_D3}. While $l_{\rm cor}$ decreases systematically with increasing traceback time, indicating that the cluster distribution was more compact at birth and has since dispersed, $D_3$ shows only a slight decrease for individual families and no systematic evolution for the young cluster sample. The young cluster sample 
also exhibits a larger $l_{\rm cor}$ than individual families, reflecting its greater spatial extent as a mixture of multiple dynamically distinct structures.
Each sample was integrated backward in time to when the youngest members were born, which means all other members were already born. The $D_3$ is therefore measured from this epoch to the present day.

\end{appendix}
\end{CJK*}
\end{document}